\documentclass[10pt,letterpaper]{article}
\usepackage[margin=1in,nohead,centering]{geometry}
\usepackage{graphicx,color,wrapfig}
\usepackage{amssymb}
\usepackage{amsmath}
\usepackage{url}
\usepackage{xspace}
\usepackage{float}
\usepackage{subfigure}
\usepackage{graphics}

\newcommand{\seq}{{\bf s}\xspace}

\newcommand{\Hermes}{\mbox{\tt Hermes}\xspace}
\newcommand{\FFTmfpt}{\mbox{\tt FFTmfpt}\xspace}
\newcommand{\FFTeq}{\mbox{\tt FFTeq}\xspace}
\newcommand{\FFTbor}{\mbox{\tt FFTbor}\xspace}
\newcommand{\FFTborTwoD}{\mbox{\tt FFTbor2D}\xspace}
\newcommand{\RNAmfpt}{\mbox{\tt RNAmfpt}\xspace}
\newcommand{\RNAeq}{\mbox{\tt RNAeq}\xspace}

\newcommand{\RNAsubopt}{\mbox{\tt RNAsubopt}\xspace}

\newcommand{\mathbbP}{\mbox{\bf P}}
\newcommand{\mathbbt}{\mbox{\bf t}}

\newtheorem{theorem}{Theorem}

\newtheorem{algorithm}[theorem]{Algorithm}

\chardef\other=12
\def\mdeactivate{
\catcode`\&=\other   \catcode`\#=\other
\catcode`\%=\other   \catcode`\~=\other
}
\def\mmakeactive#1{\catcode`#1=\active\ignorespaces}

{% The group delimits the text over which ^^M is active.
\mmakeactive\
\gdef\obeywhitespace{%
  \mmakeactive\^^M %
  \let^^M=\NewLine %
  \aftergroup\removebox %
  \obeyspaces %
}}

\def\NewLine{\par\indent}
\def\removebox{\setbox0=\lastbox}

\def\mverbatim{\par\begingroup\parindent=0em\tt\mdeactivate\obeywhitespace
\catcode`\|=0  %
}

\def\|{|}
\begin{document}

\title{Fast, approximate kinetics of RNA folding}
\author{E. Senter \and  P. Clote}
\date{}

\maketitle

\begin{abstract}
In this paper, we introduce the software suite, {\tt Hermes}, which
provides fast, novel algorithms for RNA secondary structure kinetics.
Using the fast Fourier transform to efficiently compute the Boltzmann
probability that a secondary structure $S$ of a given RNA sequence has
base pair distance $x$ [resp. $y$] from reference structure $A$ [resp.
$B$], {\tt Hermes} computes the exact kinetics of folding from $A$ to
$B$ in this coarse-grained model. In particular, {\tt Hermes} computes
the mean first passage time from the transition probability matrix by
using matrix inversion, and also computes the equilibrium time from the
rate matrix by using spectral decomposition. 
Due to the model granularity and the
speed of {\tt Hermes}, it is capable of determining secondary structure
{\em refolding kinetics} for large RNA sequences, beyond the range of 
other methods.  Comparative benchmarking of {\tt Hermes} with other
methods indicates that 
%despite the model granularity, 
{\tt Hermes} provides
refolding kinetics of accuracy suitable for use in
computational design of RNA, an important area of synthetic
biology. Source code and documentation for {\tt Hermes} are available
at \url{http://bioinformatics.bc.edu/clotelab/Hermes/}.
\end{abstract}

\section{Introduction} \label{section:introduction}

Remarkable results in RNA synthetic biology have recently been
obtained by various groups. In \cite{Hochrein.jacs13}, small
conditional RNAs have been engineered to silence a gene Y by using the
RNA interference machinery, only if a gene X is transcribed. In
\cite{Wachsmuth.nar13} a novel theophylline riboswitch has been
computationally designed to transcriptionally regulate a gene in {\em
E. coli}, and in \cite{syntheticHammerheads} a purely computational
approach was used to design functionally active hammerhead ribozymes.
Computational design of synthetic RNA molecules invariably uses some
form of thermodynamics-based algorithm; indeed, {\tt NUPACK-Design}
\cite{Zadeh.jcc11} was used to design small conditional RNAs
\cite{Hochrein.jacs13}, Vienna RNA Package \cite{Gruber08} was used in
the design of the synthetic theophylline riboswitch, and the {\tt
RNAiFold} inverse folding software
\cite{garcia.JBCBB13,GarciaMartin.nar13} was used to design the
synthetic hammerhead ribozymes. The next step in the computational
design of synthetic RNA molecules is to control the kinetics of
folding -- such control could be important in engineering
conformational switches. Kinetics is already used as a design feature
for synthetic design in proteins \cite{Bujotzek.jcam11,Fasting.acie12}.

In this paper, we introduce a new software suite, called {\tt Hermes},
which efficiently computes RNA secondary structure
folding kinetics by using a coarse-grained
method to model RNA transitions that add or remove a single base pair.
Since our motivation in developing {\tt Hermes} is to provide a new
tool to aid in engineering synthetic RNA molecules with desired
kinetic properties, {\tt Hermes} does not model {\em co-transcriptional
folding}, but only the {\em refolding} of RNA sequences.

There is a long history of both experimental and computational work on
RNA folding pathways and kinetics, so much so, that a review of
previous work is beyond the scope of this paper. Here we cite only the
most relevant related work on RNA secondary structure 
kinetics. Experimental approaches to
determine the kinetics of RNA folding include temperature jump
experiments \cite{lecuyerCrothers}, using fluorophores
\cite{Hobartner.jmb03}, using mechanical tension at the single
molecule level \cite{Vieregg.mp06}, etc. and will not be further
discussed. In this paper we focus exclusively on computational
approaches to folding kinetics, where stepwise transitions between
secondary structures involve the addition or removal of a single base
pair, as first considered in \cite{flammHofacker}. 
Nevertheless, it should be noted that there are a number of
methods that concern the addition or removal of an entire helical
region, e.g. \cite{Huang.bb14,Zhao.jcp11}.

Most computational approaches involve either (1) algorithms to
determine optimal or near-optimal folding pathways, (2) explicit
solutions of the master equation, (3) repeated simulations to fold an
initially empty secondary structure to the target minimum free energy
(MFE) structure. Examples of methods to determine optimal or
near-optimal folding pathways include the greedy approach of Morgan
and Higgs \cite{morganHiggsBarrier}, the exact, optimal, exponential
time program {\tt barriers} \cite{flammHofacker}, the program {\tt
Findpath} \cite{Flamm.r01}, which uses bounded look-ahead
breadth-first search, a genetic algorithm \cite{Shapiro.jmb01}, the
program {\tt RNAtabupath} that uses local search to determine
near-optimal folding pathways, etc. The program {\tt barriers}
\cite{flammHofacker} is the only current method guaranteed to 
produce optimal folding pathways. Since it has been shown that
the problem of determining optimal folding pathways is
NP-complete \cite{Thachuk.psb10}, it is now understandable why
{\tt barriers} can take exponential time to converge, depending
on the RNA sequence. For this reason, near-optimal solutions 
provided by heuristic methods, such as {\tt RNAtabupath}, are very useful.

Methods that employ the master equation include {\tt Treekin}
\cite{wolfingerStadler:kinetics}, which uses the programs {\tt
RNAsubopt} \cite{wuchtyFontanaHofackerSchuster} and {\tt barriers}
\cite{flammHofacker} to determine {\em macrostates}, defined as basins
of attraction near a locally optimal structure. The resulting
coarse-grained Markov chain is then sufficiently small to allow explicit
solution of the master equation. In \cite{Tang.jcb05}, a moderate
number of RNA structures were sampled according to different
strategies, from which a robotic motion planning graph was defined
to connect each sampled structure to its $k$ nearest sampled
neighbors. Again, the resulting coarse-grained Markov chain is
sufficiently small for an explicit solution of the master equation to
be given.

We now come to simulation approaches to estimate RNA folding kinetics.
The program {\tt Kinfold} \cite{flammPhD,flamm} is an implementation
of Gillespie's algorithm \cite{gillespieStochasticSimulation1},
directly related to the master equation, hence is considered by many
to be the gold standard for RNA kinetics. A recent extension of the
{\tt Kinfold} algorithm was reported in \cite{Aviram.amb12}. {\tt
Kinefold} \cite{Xayaphoummine.nar05} uses stochastic simulations of
the nucleation and dissociation of helical regions to predict
secondary structure and folding pathways. In contrast to the
previously mentioned methods, both {\tt RNAKinetics}
\cite{Danilova.jbcb06} and {\tt Kinwalker} \cite{Geis.jmb08} model
co-transcriptional folding, known to be necessary when simulating
{\em in vivo} folding of long RNA molecules \cite{Lai.r13}. As well, 
{\tt Kinefold} can simulate both refolding and co-transcriptional folding
pathways. Finally, unlike all the previous simulation results, which
depend on thermodynamic free energy parameters \cite{Turner.nar10},
the program {\tt Oxfold} \cite{Anderson.b13} performs kinetic folding
of RNA using stochastic context-free grammars and evolutionary
information.

In contrast to the previous methods, {\tt Hermes} computes the mean
first passage time (MFPT) and {\em equilibrium time} for a
coarse-grained Markov chain consisting of the ensemble of all
secondary structures having base pair distance $x$ [resp. $y$] from
reference structures $A$ [resp. $B$] of a given RNA sequence. 
Mean first passage time (MFPT) is computed exactly by matrix inversion,
and equilibrium time is computed using spectral decomposition of the rate
matrix for the coarse-grained master equation. 
{\tt Hermes} is written in C/C++, and
makes calls to the Fast Fourier Transform implementation {\tt FFTW}
\cite{FFTW05} \url{http://www.fftw.org/}, the Gnu Scientific Library
(GSL) \url{http://www.gnu.org/software/gsl/}, and the free energy
functions from Vienna RNA Package \cite{Lorenz.amb11}. The plan of the
paper is as follows. 

Section~\ref{section:algo} gives background definitions and results
concerning Markov chains, Markov processes, mean first passage time (MFPT),
and equilibrium time (ET). This section can be skipped for those readers
who are familiar with kinetics. For sufficiently small RNA sequences, MFPT and
ET are computed by {\tt Hermes} using the methods described in this section,
and so can provide a {\em gold} and {\em platinum} standard against which
other kinetics methods can be compared. Other methods include
simulation methods such as the Monte Carlo algorithm and the Gillespie
algorithm, and computations of MFPT and ET for coarse-grained Markov
chains that partition secondary structures into disjoint {\em macrostates}.
Section~\ref{section:software} presents an overview of the software suite
{\tt Hermes}, which consists of three C/C++ software packages
({\tt FFTbor2D}, {\tt RNAmfpt}, {\tt RNAeq}) and two C-driver programs
({\tt FFTmfpt}, {\tt FFTeq}). 
Section~\ref{section:results} presents the benchmarking results,
which compare {\tt Hermes} with
existent RNA kinetics software by using a collection of 1000
random 20-mers, for which accurate mean first passage times and equilibrium
times can be exactly computed. Our objective here is to show that kinetics
obtained by using coarse-grained methods from {\tt Hermes} correlate well
with the gold and platinum standards, yet are sufficiently fast to be used
in synthetic biology applications. Slower methods having finer granularity
may subsequently of great use in detailed kinetics studies of one or 
a small number of sequences.
Section~\ref{section:discussion} summarizes the contribution
of {\tt Hermes}, and the Appendix presents detailed descriptions of
all the kinetics methods and parameters used in the benchmarking results
reported in Section~\ref{section:results}.

\section{Preliminaries}
\label{section:algo}

To better understand the underlying algorithms behind the software,
we describe two traditional approaches in kinetics.

\subsection{Markov chains and mean first passage time}

Consider a physical process,
which when monitored over time, yields the stochastic sequence
$q_0,q_1,q_2,\ldots$ of discrete, observed states. If the transition from
state $q_t$ to $q_{t+1}$ depends only on $q_t$ at time
$t$ and not on the historical sequence of prior states visited, 
as often assumed in the case of protein or
RNA folding, then a Markov chain provides a reasonable
mathematical model to simulate the process.

A (first-order, time-homogeneous)
{\em Markov chain} $\mathbb{M} = (Q,\pi,P)$ is given by a finite
set $Q = \{ 1,\ldots,n \}$ of states, an initial probability distribution
$\pi = (\pi_1,\ldots,\pi_n)$, and the $n \times n$ 
{\em transition probability matrix} $P = (p_{i,j})$.
At time $t=0$, the initial state of the system is 
$q_t=i$ with probability $\pi_i$, and at discrete time 
$t=1,2,3,\ldots$, the system makes a transition from state
$i$ to state $j$ with probability $p_{i,j}$; i.e.
the conditional probability $Pr[ q_{t+1} = j | q_t=i] = p_{i,j}$.
Define the {\em population occupancy frequency} of visiting state
$i$ at time $t$ by $p_{i}(t) = Pr[q_t=i]$.  Denote
$p_{i,j}^{(t)} = Pr[q_t = j | q_0 = i]$ and notice that
the $(i,j)$-th entry of the $t$-th power $P^t$ of matrix $P$  equals
$p_{i,j}^{(t)}$.

The {\em mean first passage time} (MFPT) or
{\em hitting time} for the Markov chain $\mathbb{M}$,
starting from initial state $x_0$ and proceeding
to the target state $x_{\infty}$, is defined
as the sum, taken over all paths $\rho$ 
from $x_0$ to $x_{\infty}$, of the path length $len(\rho)$ times the probability
of path $\rho$, where $len(\rho_0,\ldots,\rho_n)$ is defined to
be $n$. In other words, $MFPT = \sum_{\rho} Pr[\rho] \cdot len(\rho)$, where
the sum is taken over sequences $\rho=\rho_0,\ldots,\rho_n$ of states
where $\rho_0=x_0$ and $\rho_n=x_{\infty}$, and $x_{\infty}$ does not
appear in $\rho_0,\ldots,\rho_{n-1}$..
%no state is visited more than once in the path $\rho$. 
%+++

Given the target state $x_{\infty}$, MFPT can be exactly 
determined by computing the inverse
$(I - P^{-}_{x_{\infty}})^{-1}$, where $I$ is the $(n-1)\times (n-1)$ identity
matrix and $P^{-}_{x_{\infty}}$ 
denotes the $(n-1)\times (n-1)$ matrix obtained from the
Markov chain transition probability matrix $P$,
by deleting the row and column with index $x_{\infty}$. 
Letting ${\bf e}$ denote the
$(n-1) \times 1$ column vector consisting entirely of $1$'s, it can be
shown that mean first passage time from state $x_0$ to state $x_{\infty}$ 
is the $x_0$-th coordinate
of column vector 
$(I - P^{-}_{x_{\infty}})^{-1} \cdot {\bf e}$ \cite{meyerMFPT}.

The {\em stationary} probability distribution $p^* = (p^*_1,\ldots,p^*_n)$ 
is a row vector such that $p^* \cdot P = p^*$; i.e. 
$p^*_j = \sum_{i}^n p^*_i \cdot p_{i,j}$, for all $1\leq j \leq n$. 
It can be shown that the stationary probability $p^*_i$ is the limit,
as $m$ tends to infinity, of the frequency of visiting state $i$ in a 
random walk of length $m$ on Markov chain $\mathbb{M}$. 
It is well-known that the stationary distribution exists and is unique
for any finite aperiodic irreducible Markov chain \cite{cloteBackofen:book}.

The Metropolis Monte Carlo algorithm \cite{metropolis:MonteCarlo} can
be used to simulate a random walk from initial state $x_0$ to target state
$x_{\infty}$, when energies are associated with the states, as is the case in
macromolecular folding, where free energies can be determined for
protein and RNA conformations from mean field theory, quantum theory,
or experimental measurements. In such cases, a {\em move set} 
defines the set $N_x$ of conformations reachable in unit time
from conformation $x$, and the transition probability matrix
$P = (p_{x,y})$ is defined as follows:
\begin{eqnarray}
\label{eqn:transitionProb1}
p_{x,y} = \left\{ 
\begin{array}{ll}
\frac{1}{|N_x|} \cdot \min\left(1,\exp(-\frac{E(y)-E(x)}{RT})\right)
&\mbox{if $y \in N_x$}\\
1 - \sum_{u \in N_x} p_{x,u} &\mbox{if $x=y$}\\
0 &\mbox{if $y \not\in N_x$, and $x \ne y$}
\end{array} \right.
\end{eqnarray}

If $p^*_x \cdot p_{x,y} = p^*_y \cdot p_{y,x}$ holds for all distinct 
$x,y \in Q$, then
{\em detailed balance} is said to hold, or equivalently the Markov
chain $\mathbb{M}$ is said to be {\em reversible}. 
If transitional probabilities
are defined as in (\ref{eqn:transitionProb1}), and if neighborhood 
size is constant ($|N_x|=|N_y|$ for all $x,y$), then it is well-known
that the stationary probability distribution
$p^*=(p^*_1,\ldots,p^*_n)$ is the Boltzmann distribution; i.e.
\begin{eqnarray}
\label{eqn:stationaryBoltzmannProb}
p^*_i = \frac{\exp(-E(i)/RT)}{Z}
\end{eqnarray}
where $E(k)$ is the energy of conformation $k$ at temperature $T$, 
$R$ is the universal gas constant, 
$T$ is absolute temperature, and the {\em partition function}
$Z = \sum_{i=1}^n \exp(-E(i)/RT)$ is a normalization constant
\cite{waterman:book,cloteBackofen:book}. If neighborhood size is not
constant, as in the case where states are RNA secondary structures and transitions
are restricted to the addition or removal of a single base pair, then
by {\em Hasting's trick}, an equivalent Markov chain can be defined which
satisfies detailed balance -- see equation (\ref{eqn:MFPTwithHastings}).

Following Anfinsen's experimental work on the denaturation and
refolding of bovine pancreatic ribonuclease \cite{anfinsen}, 
the native conformation is assumed to be the ground state having
minimum free energy. These results justify the use of the Monte Carlo
Algorithm \ref{algo:MonteCarlo} in macromolecular kinetics and 
structure prediction.

\begin{algorithm}[Metropolis Monte Carlo algorithm] \hfill\break
\label{algo:MonteCarlo}
\mverbatim
 1. procedure Metropolis(initialState $x_0$, targetState $x_{\infty}$, maxTime $T_{\max}$)
 2. //time-driven simulation of random walk from $x_0$ to $x_{\infty}$ 
 3.   $t=0$ (time); $x=x_0$ (initial state)
 4.   while $x \ne x_{\infty}$ and $t< T_{\max}$ 
 5.     choose a random neighbor $y \in N_x$
 6.     if $E(y)<E(x)$  //greedy step
 7.       x = y         //update x
 8.     else            //Metropolis step
 9.       choose random $z \in (0,1)$
10.       if $z < \exp\left(-\frac{E(y)-E(x)}{RT}\right)$
11.         x = y      //update x
12.     t += 1  //update time regardless of whether x is modified
13.   return x
|mendverbatim
\end{algorithm}

The mean first passage time from state $x$ to state $y$ can
be approximated by repeated runs of the Monte Carlo algorithm.
In particular, \v{S}ali, Shakhnovich, and Karplus used such Monte Carlo
simulations to investigate the Levinthal paradox of how a protein
can fold to its native state within milliseconds to seconds.
By repeated Monte Carlo simulations using a protein lattice model,
\v{S}ali et al. observed that a large energy difference between 
the ground state 
and the first misfolded state appears to be correlated with fast folding.

\subsection{Markov processes and equilibrium time}

A continuous time {\em Markov process} $\mathbb{M}=(Q,\pi,P(t))$
is given by a finite set $Q= \{1,\ldots,n\}$ of states, the
initial probability distribution $\pi$, 
and the $n\times n$ matrix $P(t)=(p_{i,j}(t))$ of 
probability transition functions.\footnote{Only in the
current section does $P(t)$ denote the Markov process
transition probability matrix.  In all later sections of the paper, 
$P(t) = (P_1(t),\ldots,P_n(t))$ will denote the population occupancy
frequency function, defined in (\ref{eqn:populationOccupancyEquation}).
In Markov processes, the rate matrix $R$ is used rather than the transition
probability matrix $P(t)$, so there will be no ambiguity in later
reserving $P(t)$ to denote the population occupancy function.}
Letting $q_t$ denote the state at (continuous)
time $t$, the probability that the initial state $q_0$ at time $0$ is
$k$ is $\pi_k$, while 
\begin{eqnarray}
\label{eqn:MarkovProcess}
p_{i,j}(t) = Pr[q_{t} = j| q_0 = i] .
\end{eqnarray}
The matrix $P'(t)$ of derivatives, defined by
\begin{displaymath}
  P'(t) = \left(
\begin{array}{ccc}
\frac{d p_{1,1}}{d t}(t) & \ldots  & \frac{d p_{1,n}}{d t}(t)\\
\vdots & \ddots & \vdots\\
 \frac{d p_{n,1}}{d t}(t) & \ldots  & \frac{d p_{n,n}}{d t}(t)\\
\end{array}
\right),
\end{displaymath}
can be shown to satisfy
\begin{displaymath}
P'(t) = P(t) \cdot R
\end{displaymath}
where $R = (r_{i,j})$ is an $n \times n$ {\em rate matrix} with the
property that each diagonal entry is $-1$ times the row sum
\begin{displaymath}
r_{i,i} = - \sum_{j\ne i} r_{i,j}.
\end{displaymath}
Define the {\em population occupancy} distribution 
$p(t) = (p_1(t),\ldots,p_n(t))$ by
\begin{eqnarray}
\label{eqn:markovProcessPopulationFreq}
p_i(t) = Pr[q(t) = i] = \sum_{k=1}^n\pi_k p_{k,i}(t)
\end{eqnarray}
where $q(t)$ denotes the state of the Markov process
at (continuous) time $t$. 

In the case of macromolecular folding, where Markov process states
are molecular conformations and conformational energies are available,
it is typical to define the rate matrix
$R = (r_{x,y})$ as follows:
\begin{eqnarray}
\label{eqn:transitionProb2}
r_{x,y} = \left\{ 
\begin{array}{ll}
\min\left(1,\exp(-\frac{E(y)-E(x)}{RT})\right)
&\mbox{if $y \in N_x$}\\
- \sum_{u \in N_x} p_{x,u} &\mbox{if $x=y$}\\
0 &\mbox{if $y \not\in N_x$, and $x \ne y$}.
\end{array} \right.
\end{eqnarray}
The {\em master equation} is defined by the matrix differential equation
\begin{eqnarray}
\label{eqn:masterEquationMatrix}
\frac{d p(t)}{dt} = p(t) \cdot R
\end{eqnarray}
or equivalently, for all $1 \leq x \leq n$,
\begin{eqnarray}
\label{eqn:masterEquation}
\frac{d p_x(t)}{dt} = \sum_{y=1}^n (p_y(t) \cdot r_{y,x} - p_x(t) \cdot r_{x,y})
  = \sum_{x \ne y} (p_y(t) \cdot r_{y,x} - p_x(t) \cdot r_{x,y}) .
\end{eqnarray}

As in the case of Markov chains,
$p^* = (p^*_1,\ldots,p^*_n)$ is defined to be
the stationary distribution if
$p^* \cdot P(t)= p^*$; i.e.  $p^*_k = Pr[ q(0) = k ]$ implies that
$Pr[ q(t) = k] = p^*_k$ for all $t \in \mathbb{R}$ and
$1 \leq k \leq n$. 
Define the {\em equilibrium distribution} $p^* = (p^*_1,\ldots,p^*_n)$
to be the unique solution for $p(t)$, when the master equation 
(\ref{eqn:masterEquation}) is set to equal zero; i.e.
\begin{eqnarray}
\label{eqn:equilibriumTime}
\sum_{x \ne y} p^*_x \cdot r_{x,y} = \sum_{x \ne y} p^*_y \cdot r_{y,x}.
\end{eqnarray}

If the equilibrium distribution exists, then necessarily it is equal to
the stationary distribution. A Markov process is said
to satisfy detailed balance if $p^*_x \cdot r_{x,y} = p^*_y \cdot r_{y,x}$,
for all $1 \leq x,y \leq n$, where the rate matrix $R = (r_{x,y})$.

The rate equation $R$ for is usually defined
as in (\ref{eqn:transitionProb2}) for 
Markov processes which model macromolecular folding, 
hence it is easy to see that
such Markov processes satisfy detailed balance and moreover that
the equilibrium distribution is the
Boltzmann distribution; i.e.  $p^*_x = \exp(-E(x)/RT)$ for 
all $1 \leq x \leq n$.
Since detailed balance ensures that the eigenvalues of the rate matrix $R$ are
real, one can solve the matrix differential equation
(\ref{eqn:masterEquation}) by diagonalizing the rate matrix, and thus obtain
the solution
\begin{eqnarray}
\label{eqn:populationOccupancyEquation}
p(t) = \sum_{k=1}^n c_k {\bf v}_k e^{\lambda_k t}
\end{eqnarray}
where $p(t) = (p_1(t),\ldots,p_n(t))$, and the
values $c_k$ are determined by the initial population occupancy 
distribution $p(0)$ at time $0$. Here
${\bf v}_k$ denotes the $k$th eigenvector and
$\lambda_k$ the $k$-th eigenvalue. In particular, 
$c_k = \left(p(0) \cdot T^{-1})\right)_k$, where the $j$-th row of $T$ 
is the $j$-th left eigenvector of $R$, 
and $p(0)$ is the population occupancy distribution
at time $t=0$. If the eigenvalues are labeled in decreasing
order, then $\lambda_1 \geq \lambda_2 \geq \cdots \geq \lambda_n$,
and the largest eigenvalue $\lambda_1=0$ has eigenvector $p^*$, corresponding
to the equilibrium population occupancy distribution, which in this case
is the Boltzmann distribution. The
remaining $n-1$ eigenvalues are negative, and their corresponding eigenvectors
correspond to nonequilibrium kinetic {\em relaxation modes}.

In our software {\tt Hermes}, we prefer to work with column vectors and
right eigenvectors, and so the population occupancy frequency 
$p(t)$ is defined to be the column vector
$p(t)=(p_1(t),\ldots,p_n(t))^T$. Let
$\mathbbt^1,\ldots,\mathbbt^n$ be the right eigenvectors and
$\lambda_1,\ldots,\lambda_n$ be the corresponding right eigenvalues of the
transpose $R^T$ of the rate matrix. 
Letting $T$ be the $n\times n$ matrix, whose columns are
$\mathbbt^1,\ldots,\mathbbt^n$, using standard matrix algebra
\cite{matrixTheory}, it can be shown that
\begin{eqnarray}
\label{eqn:solutionMasterEquation} 
p(t) &=& \sum_{j=1}^n (T^{-1}
\mathbbP(0))_j \cdot \mathbbt^j \cdot e^{\lambda_j t}
\end{eqnarray}
The {\em equilibrium time} can be directly computed by using a
nonlinear solver to solve for $t$ in
\begin{eqnarray}
\label{eqn:solutionEquilibriumTime}
p^* &=& \sum_{j=1}^n (T^{-1}
\mathbbP(0))_j \cdot \mathbbt^j \cdot e^{\lambda_j t}
\end{eqnarray}
where $p^*= (p^*_1,\ldots,p^*_n)^T$ and $p^*_k = \frac{\exp(-E(k)/RT)}{Z}$.
However, we have found it more
expedient to compute the {\em equilibrium time} as the smallest $t_0$,
such that for $t \in \{t_0+1,t_0+2,t_0+3,t_0+4\}$, the absolute difference
$|p(t)[x_{\infty}] - p(t_0)[x_{\infty}]| < \epsilon$, for $\epsilon =
10^{-4}$, where $x_{\infty}$ is the target RNA structure (usually taken
to be the minimum free energy structure, though this is not necessary for
the software). We also considered defining the equilibrium time to be the
smallest $t_0$, such that for $t \in \{t_0+1,t_0+2,t_0+3,t_0+4\}$, the
absolute difference $|p(t)[x] - p(t_0)[x]| < \epsilon$ for all $x \in
Q$; however, results suggest that this definition is inferior
to that just given, perhaps due to numerical instability issues when
$Q$ is taken to be the set of all secondary structures for sequences in
the benchmarking set described later.

In \cite{gillespieStochasticSimulation1} Gillespie described
a very influential algorithm to simulate a finite Markov process. 
The pseudocode, is given in Algorithm~\ref{algo:Gillespie}.
Though Gillespie's original application was for chemical kinetics, 
Flamm et al.  adapted the method for the kinetics of RNA secondary structure
folding, as implemented in {\tt Kinfold} \cite{flammPhD,flamm}.

\begin{algorithm}[Gillespie algorithm] \hfill\break
\label{algo:Gillespie}
\mverbatim
 1. procedure Gillespie(initialState $x$, targetState $w$, maxTime $T_{\max}$)
 2.   x = initial state; time $t=0$
 3.   while $x \ne w$ and $t< T_{\max}$ \{
 4.     $\Phi=0$ //$\Phi$ is flux (not probability) out of $x$
 5.     for all $y \in N_x$ do
 6.       $r_{x,y} = \min\left(1,\exp(-\frac{E(y)-E(x)}{RT})\right)$
 7.       $\Phi += r_{x,y}$
 8.     for all $y \in N_x$ do $r_{x,y} = \frac{r_{x,y}}{\Phi}$
 9.     choose random $z_1 \in (0,1)$
10.     t += $-\frac{1}{\Phi} \ln(z_1)$  //update time
11.     choose random $z_2 \in (0,1)$
12.     sum = 0
13.     for $y \in N_x$
14.       sum += $r_{x,y}$
15.       if $z_2 \leq sum$ then
16.         x=y; break //use roulette wheel
17.    \}
18.   return z 
|mendverbatim
\end{algorithm}

\section{Methods} 
\label{section:software}

The \Hermes software package was developed on the Macintosh OS X
operating system (10.9.2 and 10.10) and should work with any Unix-like platform
(Ubuntu, Debian, and CentOS were tested). We make the source code freely
available under the MIT License in two locations.
Our lab hosts the latest stable version of the code at
\url{http://bioinformatics.bc.edu/clotelab/Hermes} and a fully
version-controlled copy at \url{https://github.com/evansenter/hermes}.
The data and figures presented in this article were generated with the
source code hosted at the first URL, and we make no guarantee as to
the stability of development branches in our Git repository.

External dependencies for the software include a C (resp. C++)
compiler supporting the GNU99 language specification (resp. C++98),
FFTW implementation of Fast Fourier Transform \cite{FFTW05} ($\geq
3.3.4$) \url{http://www.fftw.org/}, Gnu Scientific Library GSL ($\geq
1.15$) \url{http://www.gnu.org/software/gsl/}, Vienna RNA Package
\cite{Lorenz.amb11} ($\geq 2.0.7$) \url{http://www.viennarna.at}, and
any corresponding sub-packages included with the aforementioned
software. For a more detailed explanation of both external
dependencies and installation instructions, refer to the 
`DOCS.pdf' file at the web site
outlining the configuration and compilation process for the \Hermes
suite.

\begin{figure*}[!th]
\begin{center}
\includegraphics[width= \textwidth]{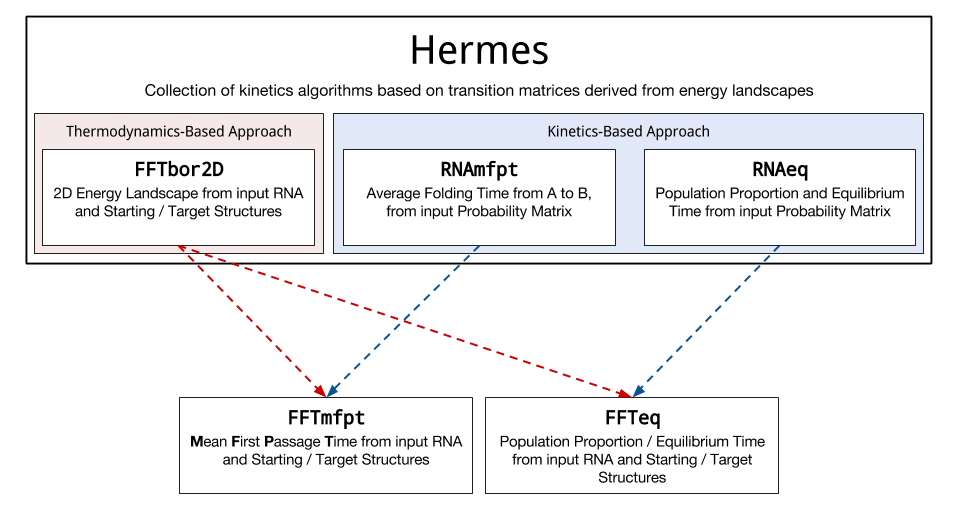} 
\caption{Overall organization of
\Hermes. \FFTborTwoD, \RNAmfpt, and \RNAeq are three distinct software
packages we have developed, which compile into both standalone
executables and archive files, providing a DRY API that allow novel
applications development using source from each of the packages,
without having to copy-and-paste relevant functions. The applications
\FFTmfpt and \FFTeq are C drivers that use data structures and functions from
\FFTborTwoD, \RNAmfpt and \RNAeq. \FFTmfpt computes the mean first
passage time (MFPT) for an RNA secondary structure to fold from an
initial structure, such as the empty structure or a given metastable
structure, into a target structure, such as the minimum free energy
(MFE) structure or possibly the Rfam \cite{Gardner.nar11} consensus
structure. \FFTeq uses spectral decomposition to compute the equilibrium time and the
fraction of the population of RNA structures that are
equal to a given target structure, as a function of time. 
 } \label{fig:organizationHermes}
\end{center}
\end{figure*}

\Hermes is organized into three independent directories: (1)
\FFTborTwoD, (2) \RNAmfpt, and (3) \RNAeq (see
Figure~\ref{fig:organizationHermes}). These packages compile into both
standalone executables and archive files. The archives provide a DRY
API which allow the development of novel applications using source
from across the \Hermes package without having to copy-and-paste
relevant functions. We provide two such examples of this in the {\tt
ext} subdirectory: \FFTmfpt and \FFTeq. These applications
are simple C drivers that use functions from \FFTborTwoD, \RNAmfpt and
\RNAeq to replicate a pipeline of executable calls without having to
deal with intermediary data transformation, I/O between calls or
slow-down due to a scripting language driver such as Python or R.

\subsection{Programs}

What follows is a brief overview of the three core applications
(\FFTborTwoD, \RNAmfpt, \RNAeq), as well as the two extension
programs (\FFTmfpt, \FFTeq). For an exhaustive list of
command-line options, please see the corresponding help files.

Given an RNA sequence $\seq$ of length $n$ and two reference
structures $A,B$, the algorithm \FFTborTwoD \cite{Senter.jmb14}
computes for each $0 \leq x,y \leq n$ the Boltzmann probability
$P(x,y) = \frac{Z(x,y)}{Z}$ of all structures having base pair
distance $x$ to $A$ [resp. $y$ to $B$]. Although the algorithm for the
software \FFTborTwoD is unchanged from that of \cite{Senter.jmb14},
the software described in this
paper has been heavily refactored to provide a user-facing API,
homogeneous with the rest of the \Hermes suite. As with the rest of
\Hermes, \FFTborTwoD makes use of either the Turner 1999 \cite{turner}
or the later Turner 2004 energy parameters
\cite{Turner.nar10}, taken respectively 
from the files rna\_turner1999.par and  rna\_turner2004.par in Vienna
RNA Package. \FFTborTwoD takes as input an RNA sequence $\seq$ and two
secondary structures $A,B$ of the input RNA sequence -- both given
in dot-bracket
notation. Alternatively, \FFTborTwoD can also process a file of RNAs,
where data for each RNA appears in the form: FASTA comment, sequence,
structure $A$, structure $B$ -- each on a separate line. In return,
for each RNA, the program returns as output a 2D probability grid 
whose value at
grid position $(x,y)$ is the Boltzmann probability of all secondary
structures compatible with $\seq$ having base pair distance $x$ (resp.
$y$) from $A$ (resp. $B$).

\RNAmfpt computes the {\em mean first passage time} (MFPT), sometimes
referred to as the {\em hitting time} of a Markov chain, by using
matrix inversion \cite{meyerMFPT} -- see
Section~\ref{section:algo}. The program takes as input a
comma separated value (CSV) file containing the non-zero positions and
values of a 2D probability grid; i.e. a CSV format file having columns {\em i},
{\em j}, and {\em p}. The first two columns, {\em i} and {\em j}
correspond to the {\em 0-indexed} row-ordered position in the rate
matrix and the final column {\em p} is the probability $p_{i,j}$ 
of a transition from {\em i} to {\em j}.  From this input, the
mean first passage time is constructed by matrix inversion.
Because this program was designed with the original intent
of handling 2D-probability grids, all vertexes are uniquely identified
by index tuples (which conceptually correspond to positioning in a 2D
array). However, it is trivial to use this code with both
1D-probability grids such as those produced by \FFTbor
\cite{Senter.po12} or arbitrary transition matrices without any change
to the underlying implementation. The software additionally provides
many options for defining the structure of the graph underlying the
Markov chain. Some of these include the option to force a fully
connected graph (useful in cases where there is no non-zero path
between the start / end state) or to enforce detailed balance.
Finally, \RNAmfpt also accepts as input the probability transition matrix,
a stochastic matrix with row sums equal to $1$, and computes the mean
first passage time for the corresponding Markov chain.

\RNAeq computes the population proportion of a user-provided structure
over arbitrary time units. Like \RNAmfpt, this program takes as input a
comma separated value (CSV) file containing the non-zero positions and
values of a 2D probability grid. From this input a rate matrix
is constructed for the underlying Markov process. Alternatively,
\RNAeq can accept as input an arbitrary rate matrix. Performing
spectral decomposition of the column-ordered rate matrix that
underlies the corresponding Markov process, \RNAeq computes either
the equilibrium time or population occupancy frequencies. 
Additionally, \RNAeq can call the
Vienna RNA Package program \RNAsubopt \cite{Wuchty.b99}, with a
user-specified upper bound to the energy difference with the minimum
free energy. With this option, the rate matrix is 
constructed for the Markov process, whose states consist of all
the structures returned by \RNAeq, and the equilibrium time or
population occupancy frequencies are computed. Due to the time
and memory required for this option, we do not expect it to be used except for
small sequences.

Beyond these three core applications, \Hermes includes two
additional programs located in the {\tt ext} directory, which fulfill
more specific roles and highlight the modular nature of the codebase.

\FFTmfpt approximates the mean first passage time of a given RNA
sequence folding from input structure A to B, by {\em exactly}
computing the mean first passage time from state $(0,d_0)$ to state
$(d_0,0)$ in the 2D probability grid obtained from running
\FFTborTwoD. Here, $d_0$ is the base pair distance between structures
$A,B$, and the MFPT is computed for the Markov chain, whose states are
the non-empty 2D probability grid points, and whose transition
probabilities are defined by $p_{(x,y),(x',y')} =
\frac{P(x',y')}{P(x,y)}$. As we report in this paper, given an RNA
sequence $\seq$, if $A$ is the empty structure and $B$ the MFE
structure of $\seq$, then \FFTmfpt output is well correlated with the
exact MFPT in folding the empty structure to the MFE structure, where
transitions between structures involve the addition or removal of a
single base pair.

\FFTeq allows an investigator to efficiently estimate population
kinetics for a sequence folding between two arbitrary, but fixed,
structures. The transition rate matrix underlying the Markov process
necessary for eigendecomposition is derived from the 2D-energy
landscape. Vertices in the rate matrix represent a subset of
structures compatible with the input sequence as modeled by
\FFTborTwoD, which makes the graph size more tractable than structural
sampling with {\tt RNAsubopt}, even with constraints.

\section{Benchmarking results}
\label{section:results}

\subsection{Kinetics methods}

In this section, we briefly describe the kinetics algorithms used in our
comparative study using a benchmarking set of 1,000 small RNAs. Each RNA from
the benchmarking set was folded from the initial empty structure to the
target minimum free energy structure.  Detailed
explanations of all kinetics programs and parameters are given
in the Appendix.

The kinetics methods benchmarked in Table~\ref{table:correlation}
are {\tt MFPT},
{\tt Equilibrium},
{\tt FFTmfpt},
{\tt FFTbor},
{\tt FFTeq} -- all computed using {\tt Hermes}, together with,
{\tt Kinfold} \cite{flammHofacker},
{\tt BarrierBasins} \cite{wolfingerStadler:kinetics},
and {\tt RNA2Dfold} \cite{hofacker:RNAbor2D}. The {\em exact} data {\tt MFPT} 
[resp. {\tt Equilibrium}] were computed using the {\tt Hermes} software {\tt RNAmfpt} 
[resp. {\tt RNAeq}] in the following fashion.

The gold standard, {\tt MFPT}, is obtained by computing the
mean first passage time by matrix inverse $(I - P^{-}_{x_{\infty}})^{-1}$
from the transition probability matrix $P$, defined by equation
(\ref{eqn:transitionProb1}) from the ensemble of all secondary structures
for each 20-mer in the benchmarking set described in the next section.
The platinum standard, {\tt Equilibrium}, is obtained from the
population occupancy frequencies
(\ref{eqn:solutionEquilibriumTime}), determined by spectral decomposition
of the rate matrix, the latter defined by 
(\ref{eqn:transitionProb2}) from the ensemble of all secondary structures
for each 20-mer in the benchmarking set.

{\tt FFTmfpt} computes the mean first passage time 
for the coarse-grained model with macrostates $M(x,y)$, consisting of
all secondary structures having base pair distance $x$ from reference
structure $A$ and distance $y$ from reference structure $B$.
{\tt FFTbor2D} is used in {\tt Hermes} to compute the Boltzmann probabilities
$p(x,y) = \frac{Z(x,y)}{Z}$ for the coarse-grained model consisting of
macrostates $M(x,y)$. Transition probabilities are computed by
equation (\ref{eqn:transitionProb1}), and matrix inversion is used to
determine mean first passage time for the coarse-grained model.
For comparison purposes, the method {\tt FFTbor} calls the
program {\tt FFTbor} \cite{Senter.po12} to compute the Boltzmann probabilities
$p(x) = \frac{Z(x)}{Z}$ for the coarse-grained model consisting of
macrostates $M(x)$ of all secondary structures having base pair distance
$x$ to a given target structure. Mean first passage time is then computed
for this model, which is even more coarse-grained than that of {\tt FFTmfpt}.
{\tt FFTeq} computes the equilibrium time
for the coarse-grained model with macrostates $M(x,y)$, consisting of
all secondary structures having base pair distance $x$ from reference
structure $A$ and distance $y$ from reference structure $B$. As in
{\tt FFTmfpt}, the program
{\tt FFTbor2D} is used to compute the Boltzmann probabilities
$p(x,y) = \frac{Z(x,y)}{Z}$ for the coarse-grained model consisting of
macrostates $M(x,y)$. The rate matrix is computed by
equation (\ref{eqn:transitionProb2}), then population occupancy
frequencies and equilibrium time
are computed by equation (\ref{eqn:markovProcessPopulationFreq})
using spectral decomposition of the rate matrix.

\begin{table*}
\begin{tabular}
  {|l|llllllll|} \hline \small{Hastings (Yes\textbackslash No)} & \small{MFPT} & \small{Equilibrium} & \small{Kinfold} & \small{FFTmfpt} & \small{RNA2Dfold} & \small{FFTbor} & \small{BarrierBasins} & \small{FFTeq} \\
  \hline \small{MFPT}   & 1      & 0.5683 & 0.7945 & 0.5060 & 0.5110 & 0.5204 & 0.5280 & 0.4472 \\
  \small{Equilibrium}   & 0.5798 & 1      & 0.7814 & 0.7043 & 0.7025 & 0.5080 & 0.5979 & 0.6820 \\
  \small{Kinfold}       & 0.7933 & 0.7507 & 1      & 0.7312 & 0.7358 & 0.6241 & 0.6328 & 0.6445 \\
  \small{FFTmfpt}       & 0.6035 & 0.7935 & 0.7608 & 1      & 0.9980 & 0.5485 & 0.8614 & 0.9589 \\
  \small{RNA2Dfold}     & 0.6076 & 0.7919 & 0.7655 & 0.9983 & 1      & 0.5584 & 0.8538 & 0.9515 \\
  \small{FFTbor}        & 0.5416 & 0.5218 & 0.6241 & 0.5748 & 0.5855 & 1      & 0.3450 & 0.4229 \\
  \small{BarrierBasins} & 0.6346 & 0.6578 & 0.6328 & 0.8310 & 0.8217 & 0.3450 & 1      & 0.9149 \\
  \small{FFTeq}         & 0.5614 & 0.7916 & 0.6966 & 0.9670 & 0.9590 & 0.4757 & 0.8940 & 1      \\
  \hline
 \end{tabular}

 \caption{Table of Pearson correlation coefficients for various methods to compute or approximate RNA secondary structure folding kinetics. Lower [resp. upper] triangular entries are with [resp. without] the Hastings correction for Markov chain probability matrices. The methods are: {\tt MFPT} (mean first passage time, computed by matrix inversion for the Markov chain consisting of all secondary structures, with move allowed between structures differing by one base pair), {\tt Equilibrium} (equilibrium time, computed by spectral decomposition of a rate matrix comprising all secondary structures to compute population fraction $P(t)$ at time $t$), {\tt Kinfold} (an implementation of Gillespie's Algorithm to approximate refolding pathways using an event-based Monte Carlo simulation), {\tt FFTmfpt} (mean first passage time for Markov chain consisting of ``grid point'' states $(x,y)$ with probability $P(x,y)=\sum_S exp(-E(S)/RT)/Z$, computed by {\tt FFTbor2D}, where the sum is taken over structures having base pair distance $x$ to the empty structure and $y$ to the MFE structure, {\tt RNA2Dfold} (mean first passage time, computed as previously explained, but using {\tt RNA2Dfold} in place of {\tt FFTbor2D} to compute $P(x,y)$), {\tt FFTbor} (mean first passage time, computed for the Markov chain consisting of states $0,1,\ldots,n$, for which $P(x) = \sum_S \exp(-E(S)/RT)/Z$, where the sum is taken over all secondary structures whose base pair distance is $x$ from the MFE structure), {\tt BarrierBasins} (equilibrium time, computed using spectral decomposition on the Markov process consisting of ``grid point'' states output from {\tt Barriers}), and {\tt FFTeq} (equilibrium time, computed in the same fashion as {\tt BarrierBasins} using a Markov process derived from the energy landscape output by {\tt FFTbor2D}). } 
\label{table:correlation}
\end{table*}

Kinetics algorithms developed by other groups and benchmarked in our study
are the following: {\tt Kinfold} \cite{flamm}, 
{\tt BarriersBasin} \cite{flammHofacker,wolfingerStadler:kinetics}, 
{\tt Kinwalker} \cite{Geis.jmb08}. Below, we give a brief overview of these methods.

{\tt Kinfold} \cite{flamm} is an implementation of Gillespie's 
Algorithm \ref{algo:Gillespie}.
For benchmarking purposes, we determined the mean first passage time
from $10^4$ {\tt Kinfold} simulations, each with an upper bound of
$10^8$ Monte Carlo steps in order to ensure convergence of each run.
{\tt BarriersBasin} is our name for the method described by Wolfinger
et al. \cite{wolfingerStadler:kinetics}, which first runs 
{\tt RNAsubopt} \cite{flammHofacker} to obtain an energy-sorted list of
secondary structures that contain the empty structure, 
then computes locally optimal structures, saddle points and macrostate
{\em basins of attraction}  using the program {\tt barriers}
\cite{wolfingerStadler:kinetics}. Subsequently, we use {\tt Hermes} to
determine the equilibrium time for the resulting Markov process whose
rate matrix is returned by {\tt barriers}.\footnote{Since 
the rate matrix returned by {\tt barriers}
does not necessarily have the property that diagonal entries are negative
row sums, as required by equation (\ref{eqn:transitionProb2}), we modify 
diagonal entries of the {\tt barriers} rate matrix to satisfy this property.} 
\footnote{At the time we did the benchmarking, we were unaware of the
program {\tt treekin} from Vienna RNA Package, which can be combined with
the rate matrix from {\tt barriers} to determine population occupation
frequency output in a file. It is necessary for the user to subsequently
write a program to determine the equilibrium time from the output of
{\tt treekin}. Since {\tt Hermes} determines equilibrium time directly
from any input rate matrix, we used {\tt Hermes} to postprocess the
output rate matrix from {\tt barriers}.}
Since {\tt Kinwalker} \cite{Geis.jmb08} implements co-transcriptional folding,
we found that its correlation with the gold and platinum standards was
extremely poor, hence we do not further report the results from our tests
with {\tt Kinwalker}. 
Programs for the remaining kinetics methods, such as the motion planning
method of Tang et al. \cite{Tang.jmb08}, the stochastic simulation
{\tt Kinefold} \cite{Xayaphoummine.nar05}, etc.
from the introduction were either not available, or else it
appeared to be difficult/impossible to extract
mean first passage times or equilibrium times to compare with the
gold and platinum standards.

Finally, Table~\ref{table:correlation} 
also includes the method we call here by {\tt RNA2Dfold}.
Given reference structures $A,B$, the {\tt RNA2Dfold}  program
\cite{hofacker:RNAbor2D,Lorenz.amb11} computes for each integer
$x,y$, the minimum free energy structure and partition function
among all secondary structures having base pair distance $x$ to $A$
and $y$ to $B$.\footnote{{\tt RNA2Dfold} generalizes an earlier algorithm
{\tt RNAbor} \cite{Freyhult.b07,Freyhult.nar07}, which computes the
minimum free energy structure and partition function among all structures
having base pair distance $x$ to reference structure $A$.}
{\tt FFTbor2D} \cite{Senter.jmb14} is a much faster program, using the
fast Fourier transform, which
computes the partition function (but not the minimum free energy
structure) over all structures having base pair distance $x$ to $A$ and 
$y$ to $B$. By the kinetics method dubbed
{\tt RNA2Dfold}, we mean that we applied
{\tt RNA2Dfold} from Vienna RNA Package 2.1.7
\cite{hofacker:RNAbor2D,Lorenz.amb11} in order
to obtain probabilities $p(x,y) = \frac{Z(x,y)}{Z}$,
and then applied {\tt Hermes} to compute the mean first passage time for
the resulting coarse-grained Markov chain, as in {\tt FFTmfpt}.

For more detailed description of each kinetics method, please see
the Appendix.

\subsection{Benchmarking data} 

In this section, we describe a
benchmarking set of 1,000 small RNAs used to benchmark the previously
described kinetics methods in a comparative study. To ensure that mean
first passage time can be computed from
$(I - P^{-}_{x_{\infty}})^{-1} \cdot {\bf e}$ by using matrix
inversion, that spectral decomposition of the rate matrix is possible,
and to ensure that {\tt Kinfold} simulations would provide sufficient
statistics, we generated a collection of 1,000 random RNA sequences of
length 20 nt, each having expected compositional frequency of $1/4$
for A,C,G,U, and each having at most 2,500 distinct secondary
structures, such that the minimum free energy is less than or equal to
$-5.5$ kcal/mol.

\begin{figure*} [!h]
\begin{center}
\includegraphics[width=0.45\linewidth]{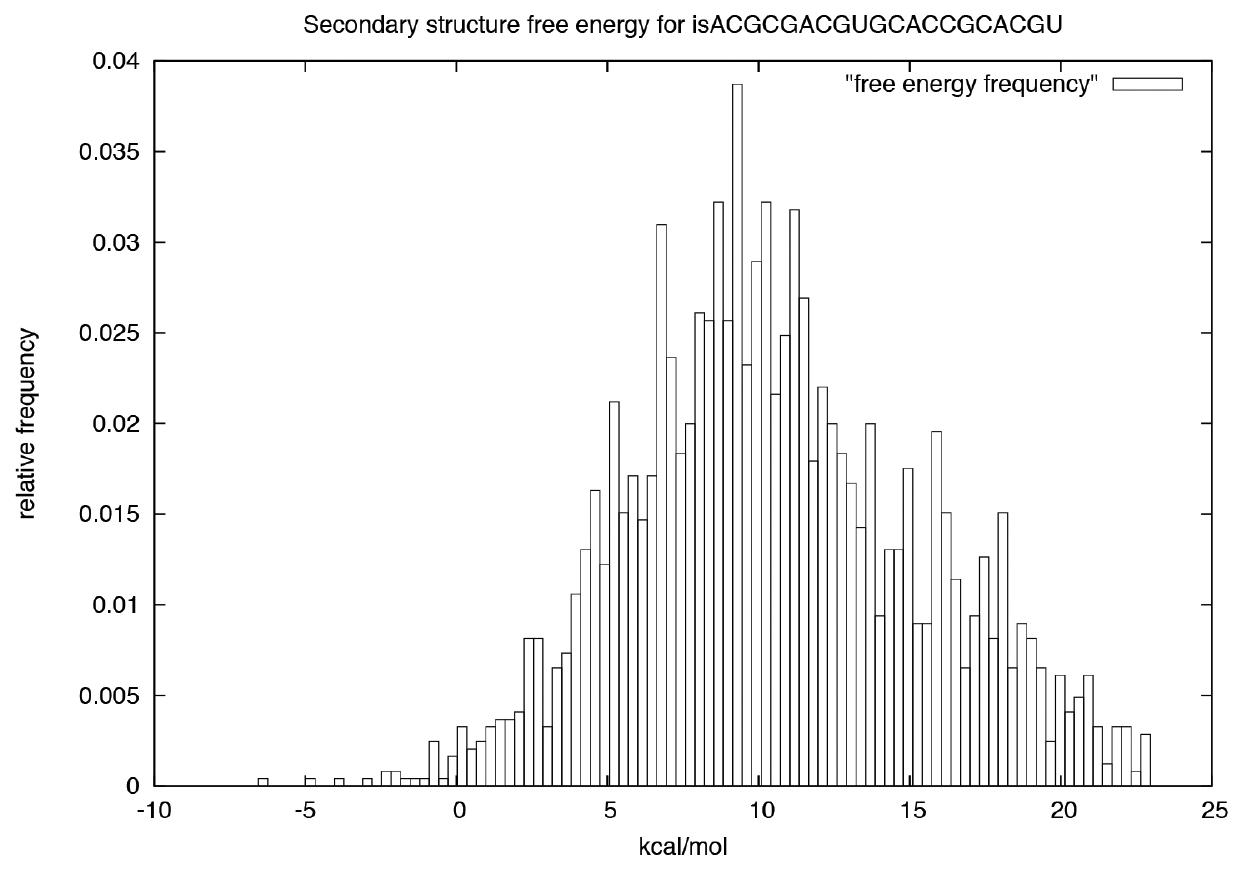}
\includegraphics[width=0.45\linewidth]{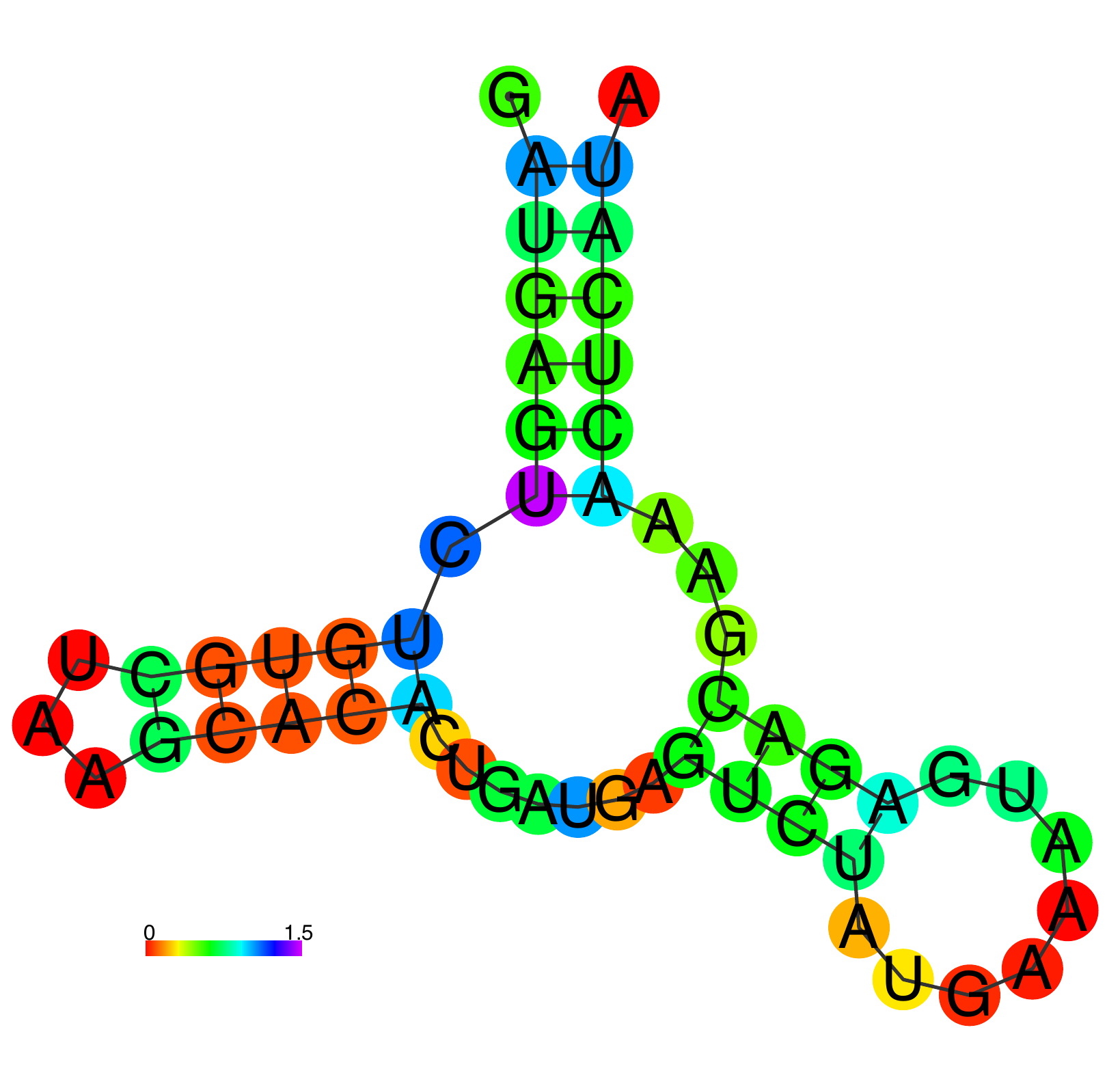}
\caption{\small
{\em (Left)} Histogram of free energies of secondary structures of
ACGCGACGUGCACCGCACGU, which range from $-6.5$ to $+25$ kcal/mol, with
mean of $10.695$ kcal/mol.
{\em (Right)} Minimum free energy structure of the 54 nt Peach Latent Mosaic 
Viroid (PLMVd) AJ005312.1/282-335, which is identical to the consensus
structure from Rfam 11.0 \cite{Gardner.nar11}. {\tt RNAfold} from
Vienna RNA Package 2.1.7 with energy parameters from the Turner 1999
model were used, since the minimum free energy structure determined by
the more recent Turner 2004 energy parameters 
does {\em not} agree with the Rfam consensus structure -- see 
\cite{syntheticHammerheads}. Positional entropy, a measure
of divergence in the base pairing status at each positions for the
low energy ensemble of structures, is indicated by color, using the
RNA Vienna Package utility script {\tt relplot.pl}.
}
\label{fig:PLMVd}
\end{center}
\end{figure*}

For example, one of the 1,000 sequences is ACGCGACGUGCACCGCACGU with
minimum free energy structure {\tt .....((((((...))))))} having free
energy of $-6.4$ kcal/mol. Statistics for the free energies of the
2,453 secondary structures of this 20-mer are the following: mean is
$10.695$, standary deviation is $4.804$, maximum is $25.00$, minimum
is $-6.40$. A histogram for the free energy of all secondary
structures of ACGCGACGUGCACCGCACGU is depicted in
%Figure~\ref{fig:freeEnergyHistogramSecStrACGCGACGUGCACCGCACGU}. 
the left panel of Figure~\ref{fig:PLMVd}. The right panel of the
same figure depicts the minimum free energy structure of the
54 nt hammerhead type III ribozyme from Peach Latent Mosaic Viroid
(PLMVd), discussed later. This secondary structure is identical
to the consensus structure from Rfam 11.0 \cite{Gardner.nar11}.

\begin{figure*}
\centering
\includegraphics[width=0.45
\textwidth]{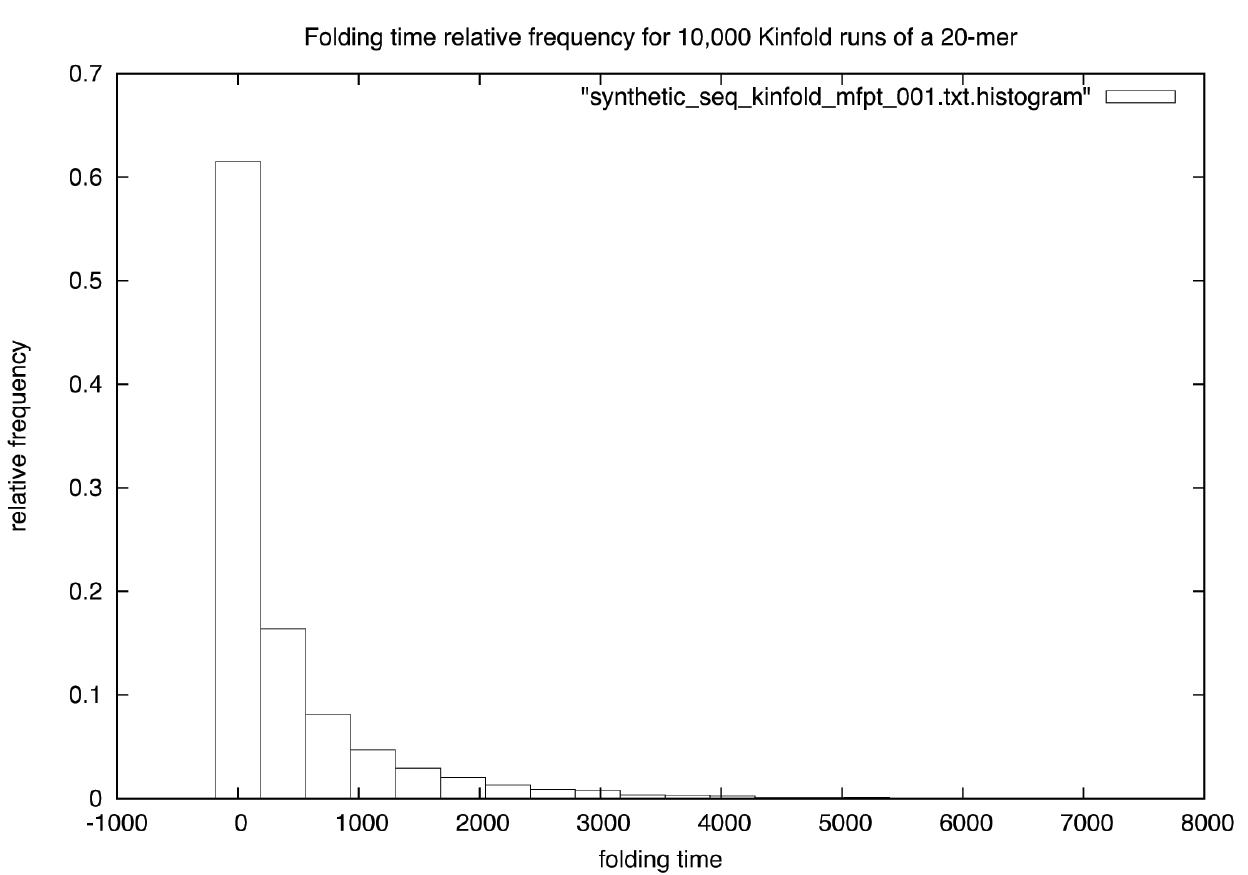}
\includegraphics[width=0.45
\textwidth]{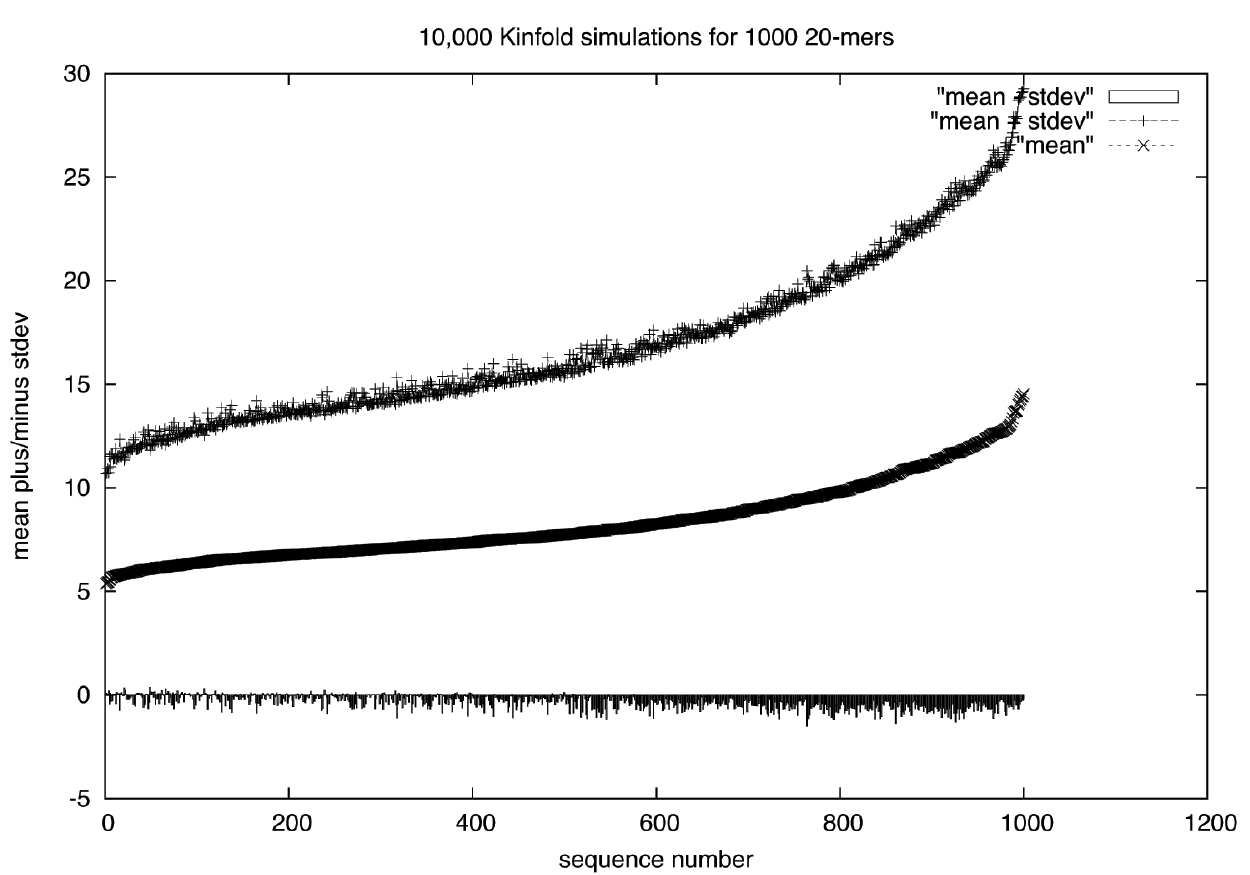} \caption{{\em (Left)} Histogram of {\tt Kinfold} folding times for 20-mer CCGAUUGGCG AAAGGCCACC. The mean [resp. standard deviation] of 10,000 runs of {\tt Kinfold} for this 20-mer is 538.37 [resp. 755.65]. Note the close fit to the exponential distribution, {\em (Right)} Mean minus standard deviation ($\mu -\sigma$), mean ($\mu$), and mean plus standard deviation ($\mu + \sigma$) of the logarithm of {\tt Kinfold} folding times, taken over 10,000 runs for each of the 1,000 sequences from the benchmarking set of 20-mers. For graphical illustration, we have sorted the log folding times in increasing order. } \label{fig:meanStdevKinfoldRuns1000sequences}
\end{figure*}

Figure~\ref{fig:meanStdevKinfoldRuns1000sequences} displays the mean
and standard deviation for {\tt Kinfold} simulations of folding time
for each of the 1,000 RNA sequences from our benchmarking data. For
each sequence, the mean and standard deviation of the time required to
fold the empty structure to the MFE structure were computed from
10,000 {\tt Kinfold} runs, each run with an upper bound of $10^8$
Monte Carlo steps, thus ensuring that all simulations converged. The
sequences were then sorted by increasing folding time mean. Standard
deviation exceeded the mean in $83.9\%$ of the 1,000 cases, indicating
the enormous variation between separate {\tt Kinfold} runs, even for
20 nt RNA sequences having at most 2,500 secondary structures. In our
opinion, {\tt Kinfold} is an expertly crafted implementation of
Gillespie's algorithm for an event driven Monte Carlo simulation of
one-step RNA secondary structure folding. From the standpoint of
biophysics and physical chemistry, there is no more reliable
simulation method, except of course the exact computation of mean
first passage time using linear algebra. Nevertheless, the enormous
time required for reliable {\tt Kinfold} estimations and the large
standard deviations observed point out the need for a faster method to
approximate folding time.

\subsection{Correlations}

In this section, we display the correlation between (1) the {\em gold
standard} method {\tt MFPT}, both with and without the Hastings
modification using equations (\ref{eqn:MFPTwithHastings}) and
(\ref{eqn:MFPTwithoutHastings}), (2) the {\em platinum standard} method {\tt
Equilibrium}, (3) the {\em silver standard} method {\tt Kinfold}, (4) {\tt
FFTmfpt} with and without the Hastings modification using equations
(\ref{eqn:transitionProbabilityFFTbor2DwithHastings}) and
(\ref{eqn:transitionProbabilityFFTbor2DwithoutHastings}), (5) {\tt
FFTeq} which computes equilibrium time for the 2D-grid, (6) {\tt
RNA2Dfold } with and without the Hastings modification using equations
(\ref{eqn:transitionProbabilityFFTbor2DwithHastings}) and
(\ref{eqn:transitionProbabilityFFTbor2DwithoutHastings}). Correlations
with [resp. without] the Hastings modification are summarized in the
lower [resp. upper] triangular portion of
Table~\ref{table:correlation}. It is clear that correlations between
the mathematically exact methods {\tt MFPT}, {\tt Equilibrium}, and
approximation methods {\tt Kinfold}, {\tt FFTmfpt}, {\tt FFTeq}, {\tt
RNA2Dfold} are improved when using the Hastings correction.

\begin{figure}
\centering
\includegraphics[width=0.45 \textwidth]{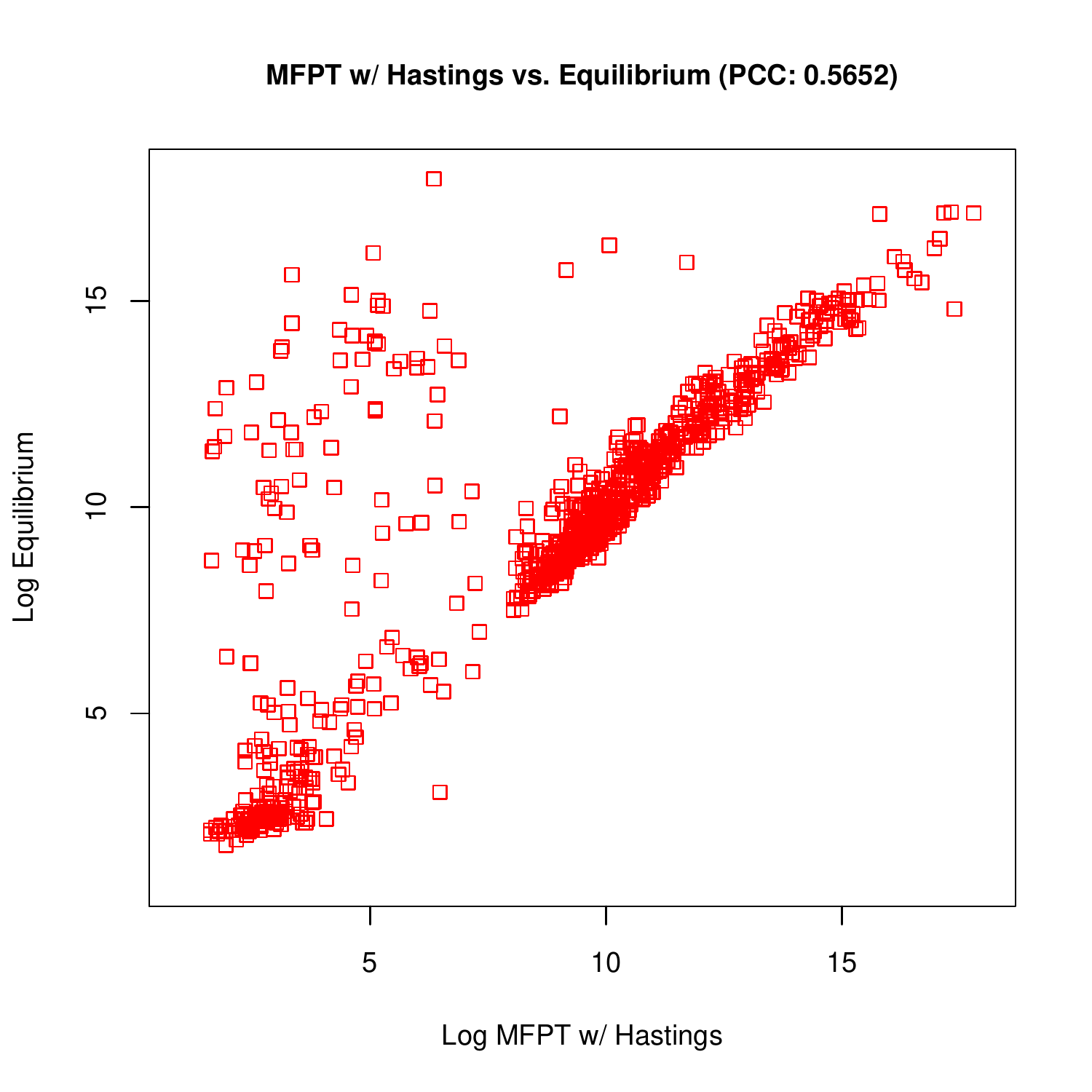}
\includegraphics[width=0.45 \textwidth]{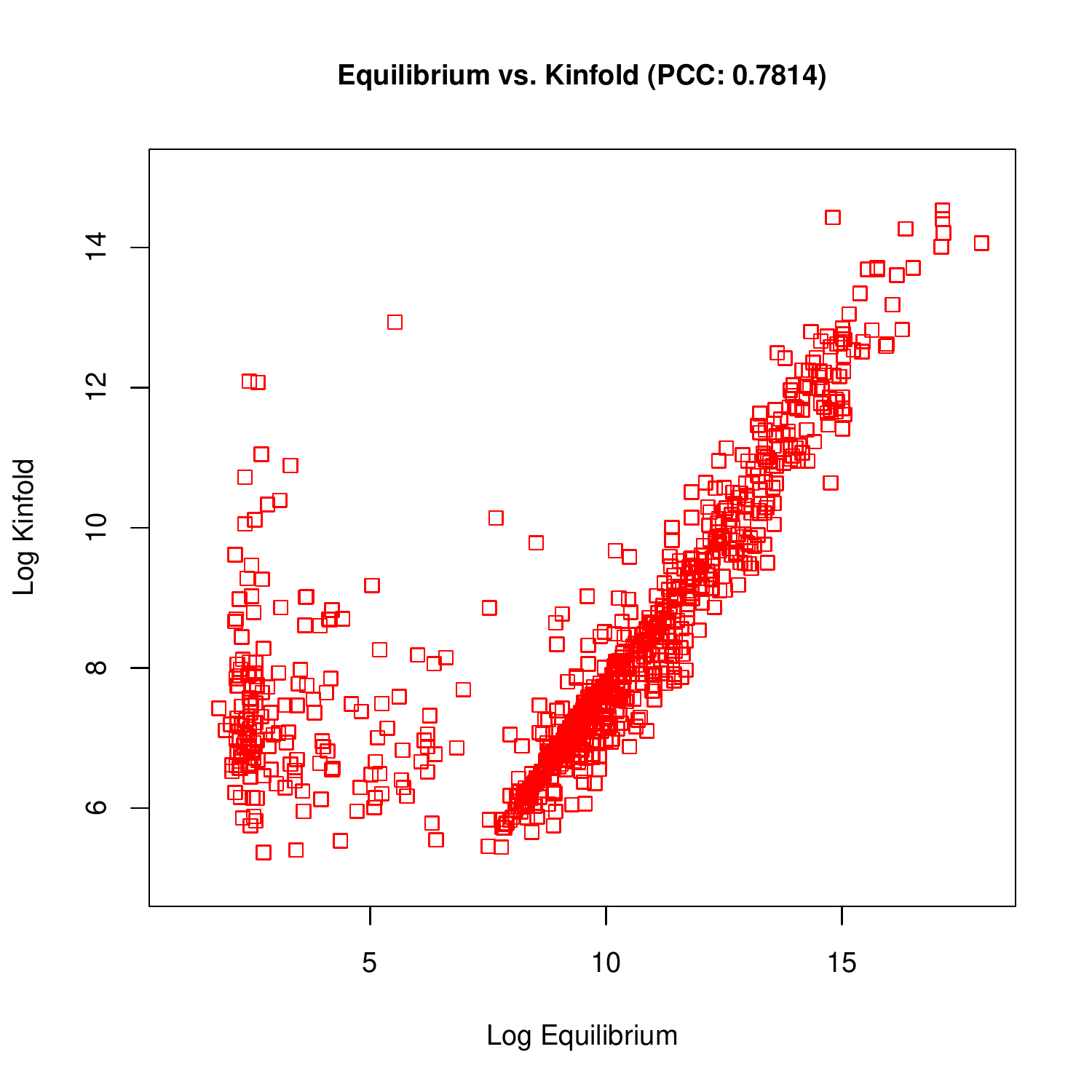}
\caption{ Scatter plots of the natural logarithm of times from {\tt
MFPT} versus {\tt Equilibrium} (left) and for {\tt Kinfold} versus
{\tt Equilibrium} (right). }
\label{fig:scatterplot_single_bp_moves_for_kinfold_vs_rnaeq}
\end{figure}

\begin{figure}
\centering
\includegraphics[width=0.45 \textwidth]{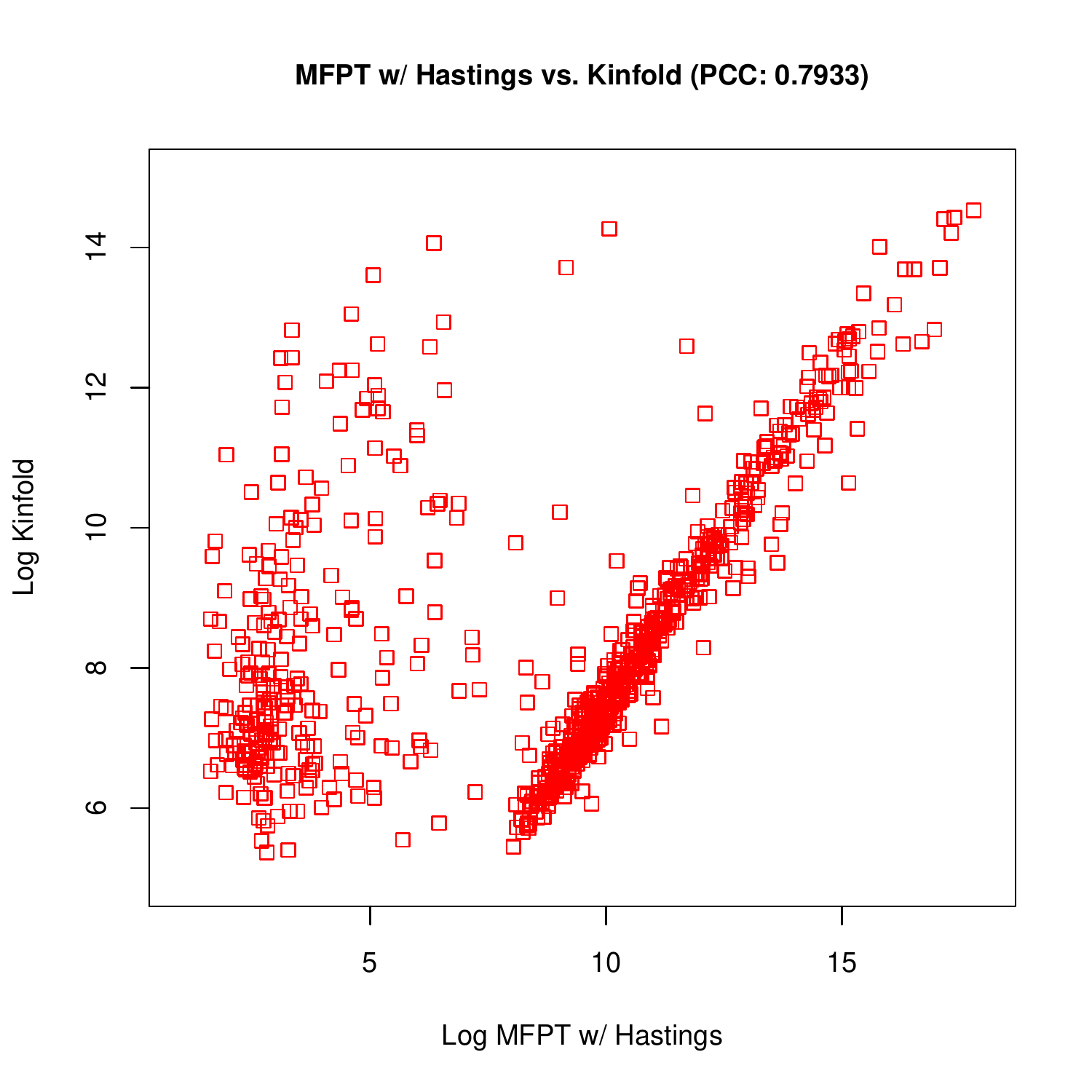}
\includegraphics[width=0.45 \textwidth]{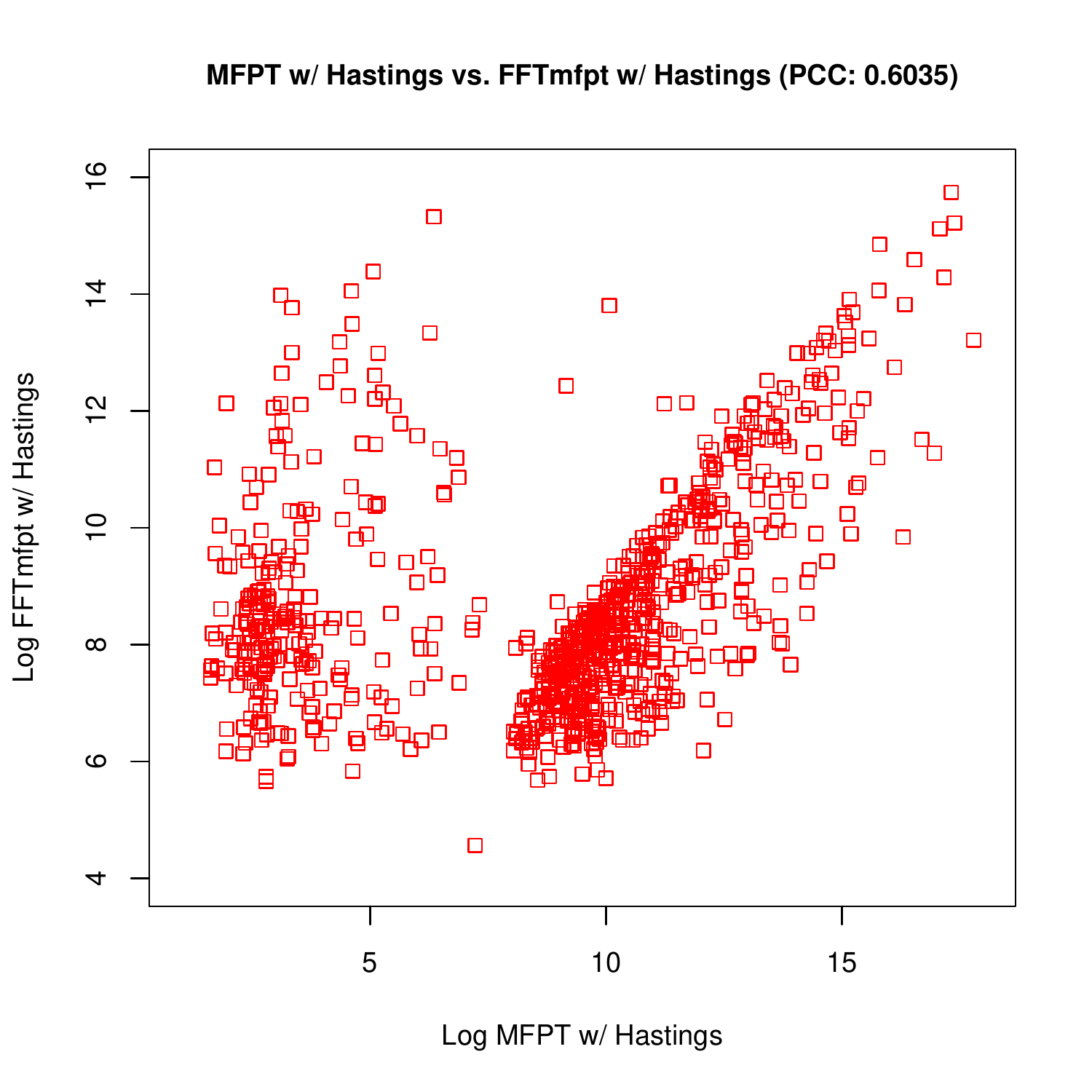}
\caption{ Scatter plots of the natural logarithm of times from {\tt
MFPT} versus {\tt Kinfold} (left) and for {\tt MFPT} versus {\tt
FFTmfpt} (right). }
\label{fig:scatterplot_single_bp_moves_for_actual_mfpt_vs_fftbor2d}
\end{figure}

\begin{figure}
\centering
\includegraphics[width=0.45 \textwidth]{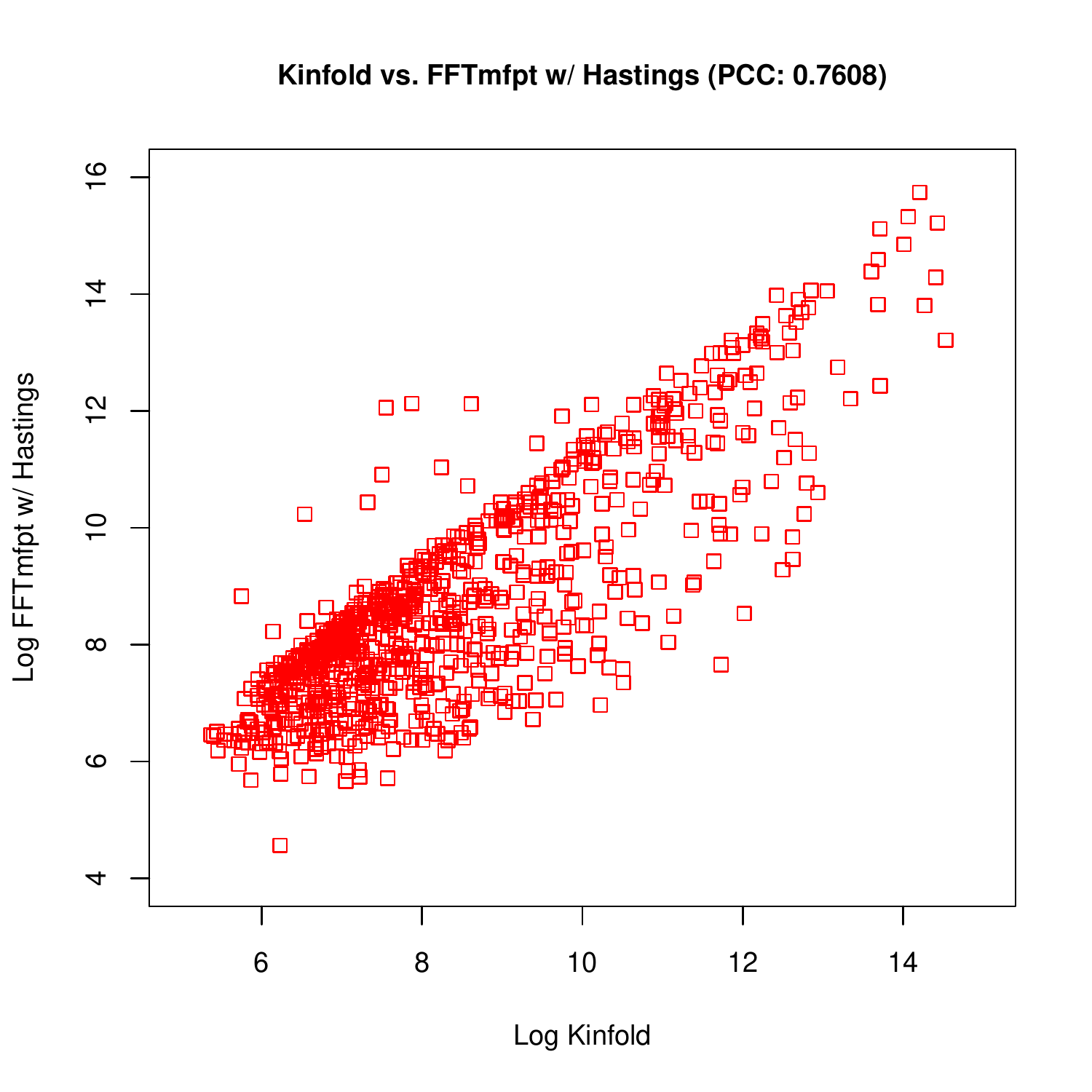}
\includegraphics[width=0.45 \textwidth]{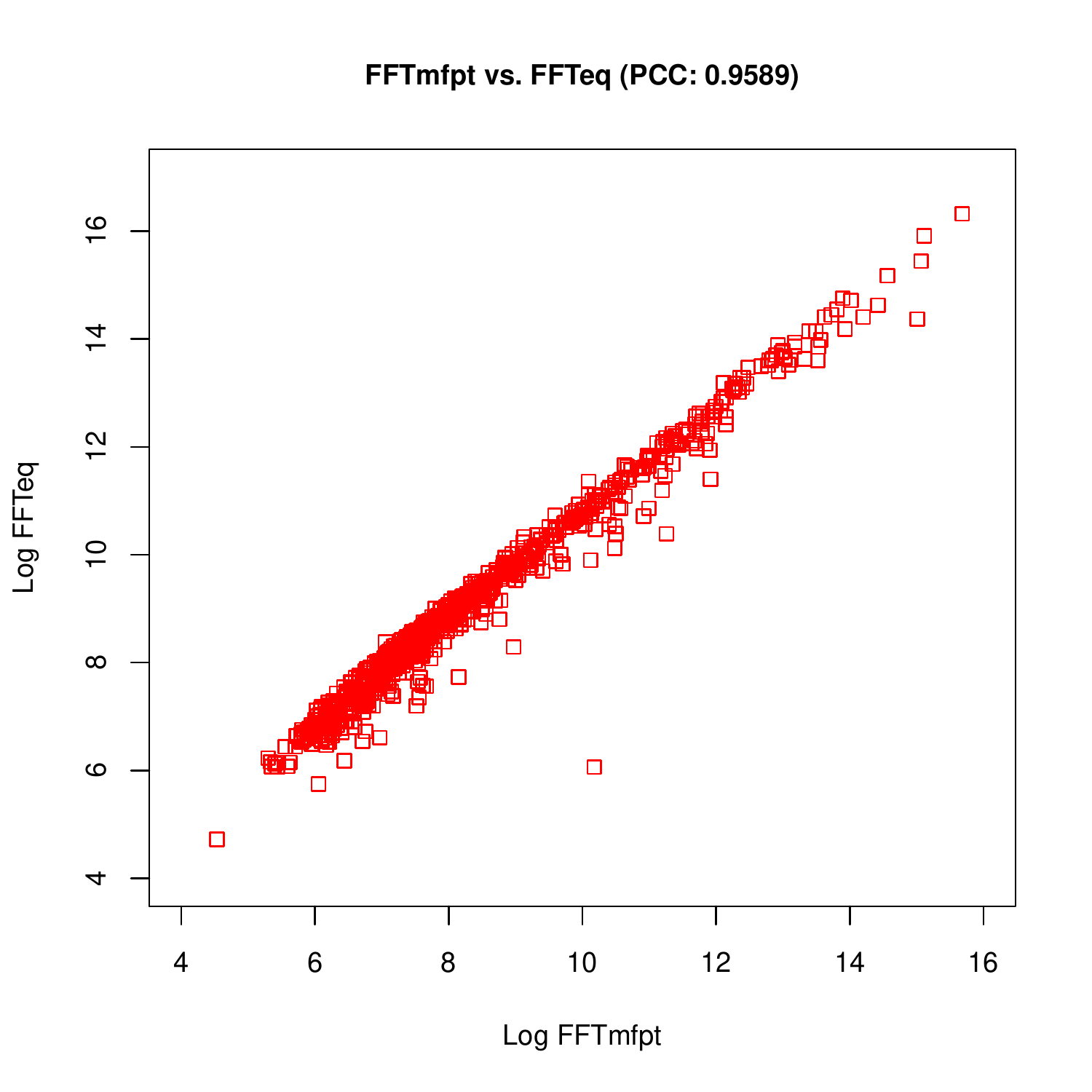}
\caption{ Scatter plots of the natural logarithm of times from {\tt
Kinfold} versus {\tt FFTmfpt} (left) and for {\tt FFTmfpt} versus {\tt
FFTeq} (right). }
\label{fig:scatterplot_single_bp_moves_for_fftbor2d_vs_rnapopulation}
\end{figure}

Figures~\ref{fig:scatterplot_single_bp_moves_for_kinfold_vs_rnaeq},
\ref{fig:scatterplot_single_bp_moves_for_actual_mfpt_vs_fftbor2d},
\ref{fig:scatterplot_single_bp_moves_for_fftbor2d_vs_rnapopulation}
depict scatterplots for kinetics obtained by some of the algorithms
above. The left panel of
Figure~\ref{fig:scatterplot_single_bp_moves_for_kinfold_vs_rnaeq}
shows a scatter plots for gold standard {\tt MFPT} versus platinum
standard {\tt Equilibrium}, with correlation value 0.5652. The right
panel of the same figure shows a scatter plot for {\tt Kinfold} versus
{\tt Equilibrium}, with correlation 0.7814. Note the persence of two
clusters in this and some of the other scatter plots. Cluster A
consists of RNA sequences whose folding time, as determined by {\tt
MFPT} or {\tt Equilibrium}, is rapid -- specifically, the natural
logarithm of the MFPT is at most 7.5. Cluster B consists of the
remaining RNA sequences, whose folding time is longer than that of
cluster A. There are no significant differences between RNA sequences
in clusters A and B with respect to GC-content, sequence logo, minimum
free energy, number of secondary structures, etc.  The left panel of
Figure~\ref{fig:scatterplot_single_bp_moves_for_actual_mfpt_vs_fftbor2d}
shows the scatter plot for {\tt MFPT} versus {\tt Kinfold}, with
correlation 0.7933, and the right panel shows the scatter plot for
{\tt MFPT} versus {\tt FFTmfpt}, with correlation 0.6035.
Figure~\ref{fig:scatterplot_single_bp_moves_for_fftbor2d_vs_rnapopulation}
shows scatter plots for {\tt FFTmfpt} versus {\tt Kinfold} (left) and
for {\tt FFTmfpt} versus {\tt FFTeq} (right), with respective
correlation values 0.7608 and 0.9589. {\tt Kinfold} obviously provides
a better correlation with the exact value of mean first passage time;
however, since the standard deviation of {\tt Kinfold} runs is as
large as the mean,\footnote{It follows from spectral decomposition that
equilibrium time follows an exponential distribution (or sum of 
exponential distributions). Exponential distributions have the property
that the mean is equal to the standard deviation, hence
it is not surprising that {\tt Kinfold}
kinetics have this property.} accurate kinetics estimates 
from {\tt Kinfold} require prohibitively large computational time -- indeed, in
\cite{wolfingerStadler:kinetics} reliable kinetics for phe-tRNA from
yeast were obtained by 9,000 {\tt Kinfold} simulations, each for $10^8$
steps, requiring 3 months of CPU time on an Intel Pentium 4 running at
2.4 GHz under Linux. Although the correlation value of 0.6035 between
{\tt MFPT} and {\tt FFTmfpt} is much less than that obtained by {\tt
Kinfold}, the runtime required by our method {\tt FFTmfpt} is measured
in seconds, even for moderate to large RNAs. For this reason, we
advocate the use of {\tt FFTmfpt} in synthetic biology screens to
design RNA sequences having certain desired kinetic properties. Once
promising candidates are found, it is possible to devote additional
computational time to {\tt Kinfold} simulations for more accurate
kinetics.

\section{Discussion} 
\label{section:discussion}

In this paper, we have introduced 
the fast, novel RNA kinetics software suite, {\tt
Hermes}, which consists of the three software packages, {\tt FFTbor2D},
{\tt RNAmfpt}, and {\tt RNAeq},  together with two C drivers,
{\tt FFTmfpt} and {\tt FFTeq}.

Given an RNA sequence $\seq$,  {\tt FFTbor2D} computes the Boltzmann probability
$p(x,y) = \frac{Z(x,y)}{Z}$ of secondary structures of $\seq$, whose base
pair distance from initial structure $A$ is $x$ and from target structure 
$B$ is $y$. {\tt FFTbor2D} is an enhancement of the algorithm 
described in \cite{Senter.jmb14}, shown there
to run both in theory and in practice much faster than {\tt RNA2Dfold}
\cite{hofacker:RNAbor2D}, 
since {\tt FFTbor2D} uses the fast Fourier transform to
compute Boltzmann probabilities $p(x,y)$ by polynomial interpolation.

\begin{figure*} [!ht]
\begin{center}
\includegraphics[width=.45\linewidth]{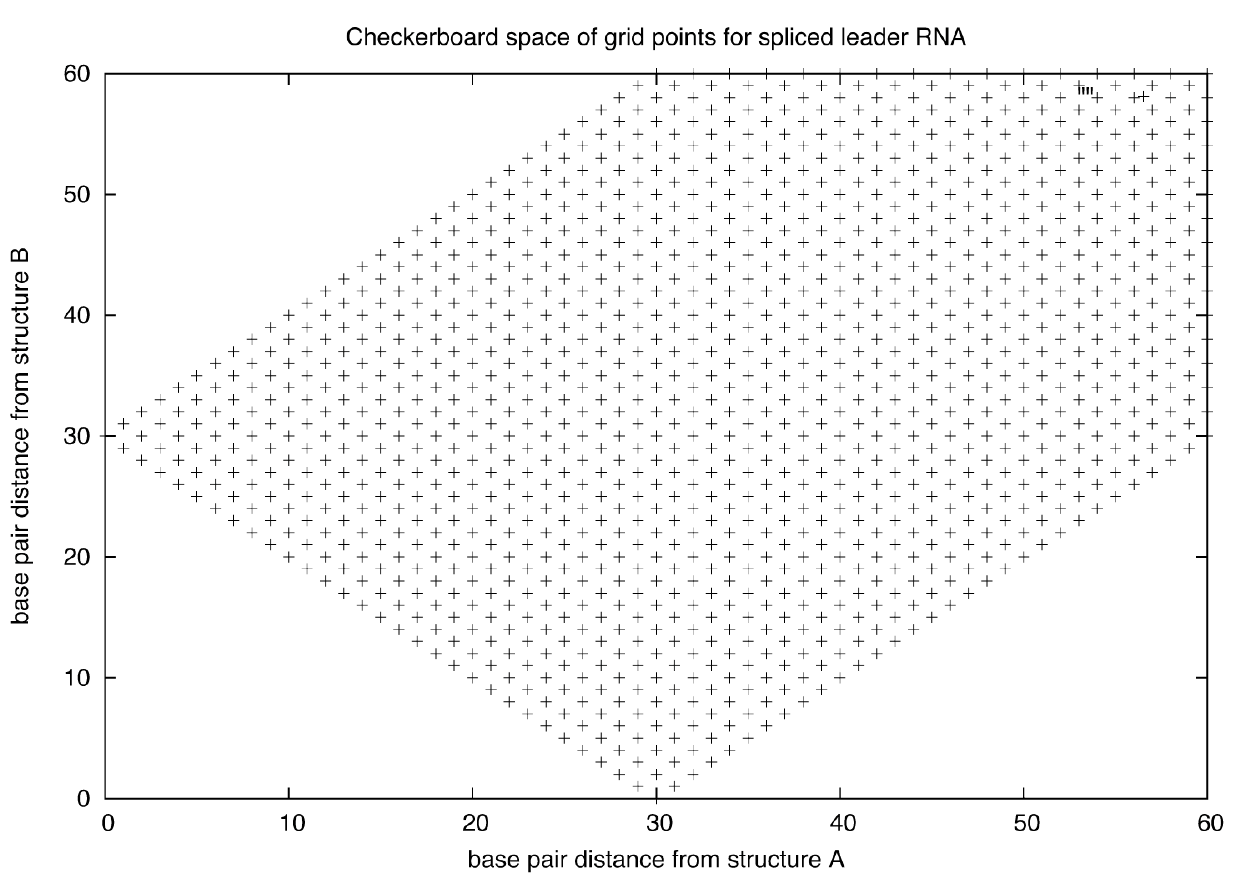}
\includegraphics[width=.45\linewidth]{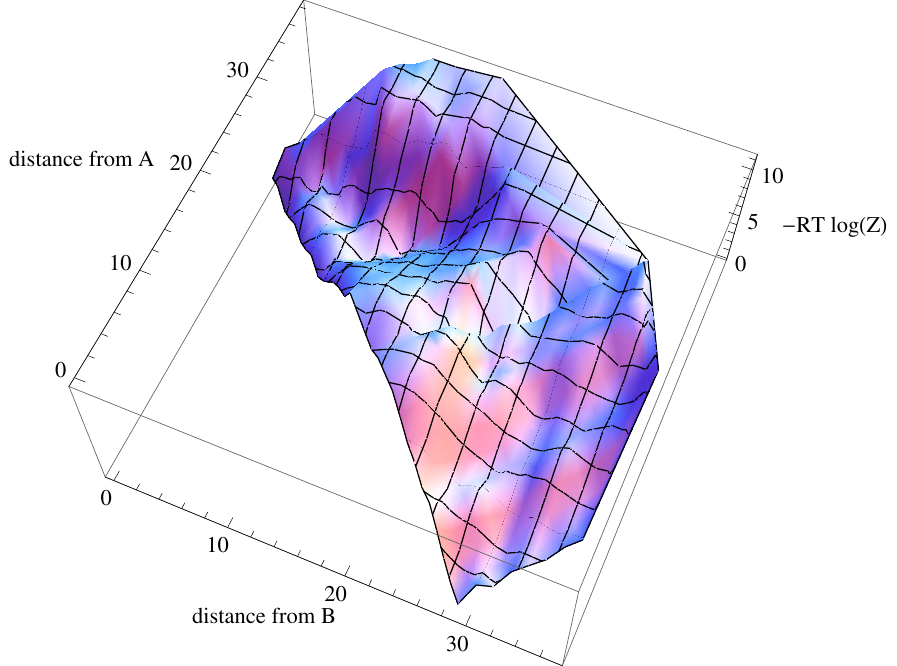}
\caption{\small {\em (Left)}
As first pointed out in \cite{hofacker:RNAbor2D}, the
set of grid points $(x,y)$, for which a secondary structure $S$ has
base pair distance $x$ from reference structure $A$ and base pair
distance $y$ from reference structure $B$, forms a kind of
checkerboard image. For the 56 nt spliced leader RNA from {\em L.
collosoma} \cite{lecuyerCrothers}, with sequence AACUAAAACA AUUUUUGAAG
AACAGUUUCU GUACUUCAUU GGUAUGUAGA GACUUC, and reference structure $A$
given by {\tt
..((...((((((..(((((.((((...)))).)))))..))).)))..)).....}, and
reference structure $B$ given by {\tt
.......................((((((((((((.....)))))..)))))))..}, we obtain
the possible grid points depicted in this figure. 
The $x$-axis (resp. $y$-axis) 
represents base pair distance between initial structure $A$ (resp. target
structure $B$).
Consideration of the triangle inequality and parity condition
(see text) ensure that partition function values $Z(x,y)$ and Boltzmann
probabilities $p(x,y) = \frac{Z(x,y)}{Z}$ are zero for all 
{\em non}-grid points $(x,y)$.
{\em (Right)}
2D projection of energy landscape for Spliced Leader (SL) RNA 
from {\em Leptomonas collosoma}, in which
the $z$-axis represents the ensemble free energy $-RT \log Z(x,y)$,
where $Z(x,y)$ is computed in {\tt FFTbor2D} by $Z_{x,y}= p(x,y) \cdot Z$.
Low energy positions $(x,y)$ correspond to high Boltzmann probability positions.
Image taken from \cite{Senter.jmb14}.
}
\end{center}
\label{fig:checkerboardGridPointsSplicedLeader}
\end{figure*}

{\tt RNAmfpt} computes the mean first passage time for the 
coarse-grained Markov chain,
consisting of the macrostates $M(x,y)$ of all secondary structures,
whose base pair distance is $x$ [resp. $y$] from initial structure $A$ 
[resp. target structure $B$]. 
Given an initial state $A=(x_0,y_0)$, a
target state $B=(x_1,y_1)$, and the 2D grid of positions $(x,y)$
and probabilities $p(x,y)$ as computed by {\tt FFTbor2D} and
illustrated in
Figure~\ref{fig:checkerboardGridPointsSplicedLeader},
{\tt RNAmfpt} computes the {\em mean first passage time}, taken
over all paths from $(x_0,y_0)$ to  $(x_1,y_1)$.  Alternatively,
{\tt RNAmfpt} can take as input the probability transition matrix $P$ for 
an arbitary finite Markov chain $\mathbb{M}$, and subsequently
determines the {\em mean first passage time} by computing the inverse 
$(I - P^{-}_{x_{\infty}})^{-1} = \displaystyle\sum_{k=0}^{\infty}
(P^{-}_{x_{\infty}})^k$.

As in the case of {\tt RNAmfpt}, {\tt RNAeq} 
computes the equilibrium time for the coarse-grained Markov process,
consisting of the macrostates $M(x,y)$ of all secondary structures,
whose base pair distance is $x$ [resp. $y$] from initial structure $A$ 
[resp. target structure $B$]. 
Given an initial state $A=(x_0,y_0)$, a
target state $B=(x_1,y_1)$, and the 2D grid of positions $(x,y)$
and probabilities $p(x,y)$ as returned by {\tt FFTbor2D} and
illustrated in
Figure~\ref{fig:checkerboardGridPointsSplicedLeader},
{\tt RNAeq} computes the population occupancy frequencies 
$P(t) = (P_1(t),\ldots,P_n(t))$ by using spectral decomposition, and
then determines the {\em equilibrium time}. 
Alternatively, given an input RNA sequence {\tt RNAeq} 
calls {\tt RNAsubopt} \cite{wuchtyFontanaHofackerSchuster} to generate
structures within a user-specified energy bound (or to sample a user-specified
number of structures from the low energy ensemble). {\tt RNAeq} subsequently
computes the rate matrix and computes population occupancy frequencies
and equilibrium time. {\tt RNAeq} can additionally
take as input the rate matrix $R$ for 
an arbitary finite Markov process $\mathbb{M}$, and then
compute population occupancy frequencies
and the {\em equilibrium time}.

Given an RNA sequence,
the C-drivers {\tt FFTmfpt} and {\tt FFTeq} respectively compute
the mean first passage time (MFPT) and
equilibrium time (ET) for the coarse-grained Markov chain on the
macrostates $M(x,y)$ of all secondary structures whose base pair
distance is $x$ [resp. $y$] from initial structure $A$ [resp. target
structure $B$]. This is done by calling {\tt FFTbor2D}, in order to
determine the 2D probabilities, from which the transition probability
matrix or respectively rate matrix are determined, and then calling
{\tt RNAmfpt} or respectively {\tt RNAeq}. 

Using functions from Gnu Scientific Library (GSL)
\url{http://www.gnu.org/software/gsl/}, mean first passage time 
(MFPT) is computed by matrix inversion
$(I - P^{-}_{x_{\infty}})^{-1}$ from the transition probability matrix
$P$, while equilibrium time (ET) is computed by spectral
decomposition of the rate matrix. {\tt Hermes} makes available a
variety of novel RNA folding kinetics methods: {\tt MFPT} (via {\tt RNAmpft}), {\tt
Equilibrium} (via {\tt RNAeq}), {\tt FFTmfpt}, and {\tt FFTeq}.
Using a collection of 1,000 randomly generated 20-mers having at most
2,500 secondary structures, using {\tt Hermes} the gold standard 
mean first passage time and platinum standard equilibrium time were
computed for each sequence.
Pearson correlation was then computed
between the folding times of all the methods of {\tt Hermes}, along
with the current state-of-the-art kinetics programs {\tt Kinfold}
\cite{flamm}, {\tt BarrierBasins} \cite{wolfingerStadler:kinetics},
{\tt Kinwalker} \cite{Geis.jmb08}. 

{\tt Kinfold} is an implementation
of the Gillespie algorithm \cite{gillespieStochasticSimulation1},
which essentially samples from an exponential distribution of folding
times, hence {\tt Kinfold} mean and standard deviation are
approximately equal -- see
Figure~\ref{fig:meanStdevKinfoldRuns1000sequences}. Elementary
considerations from statistics indicate that for our benchmarking set
of 20-mers, the minimum sample size $n = \left( \frac{z_{\alpha/2}
\cdot \sigma}{E} \right)^2$ ranges from 937,712 to 23,289,310 for us
to have a confidence level of 95\% that the average of $n$ {\tt
Kinfold} runs differs from the real folding time by at most 100 steps.
Since this is the case for tiny sequences of 20 nt, it follows that
only enormous computation time can provide reliable folding times for
larger sequences when using {\tt Kinfold}.

\begin{figure*}
\centering
\includegraphics[width=0.45 \textwidth]{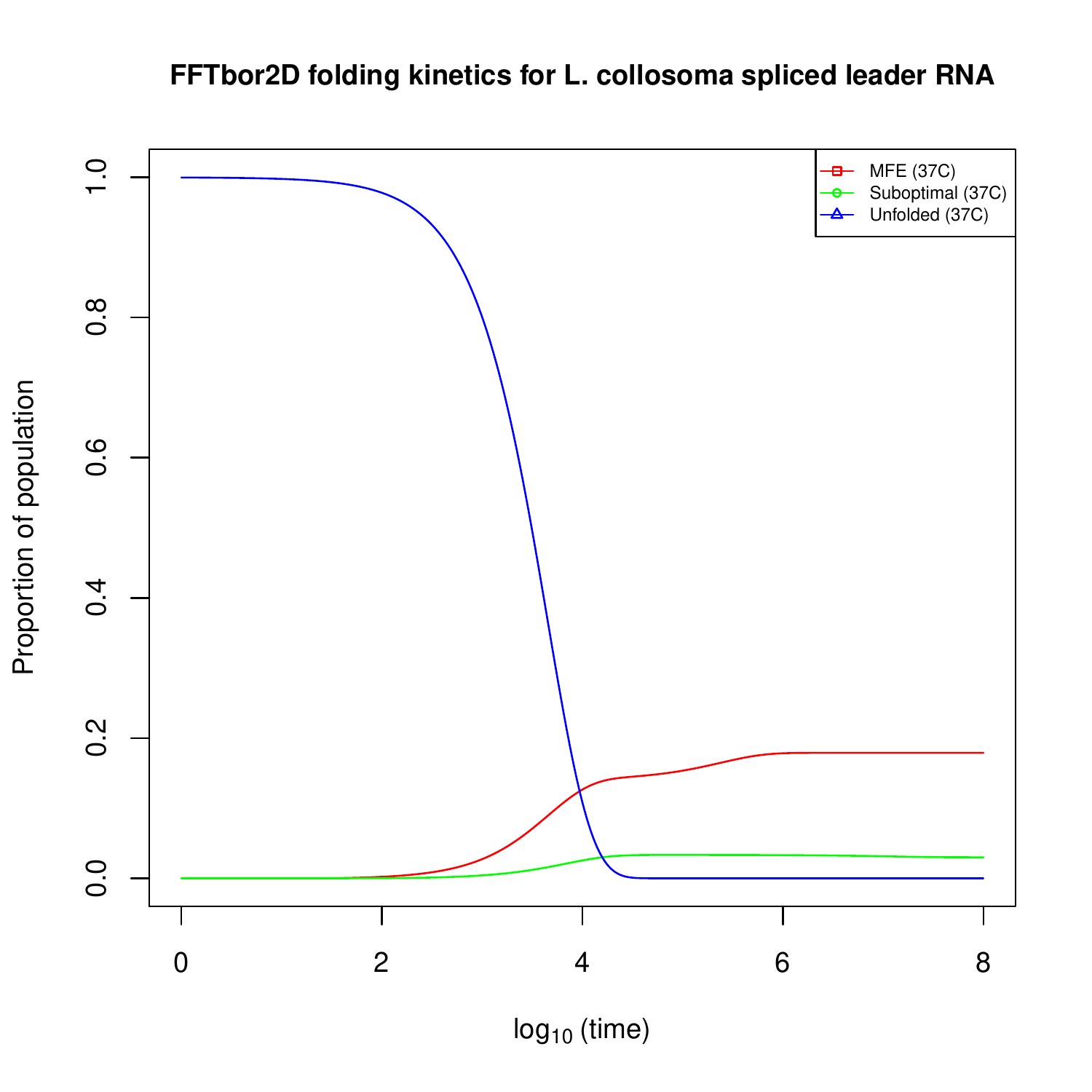}
\includegraphics[width=0.45 \textwidth]{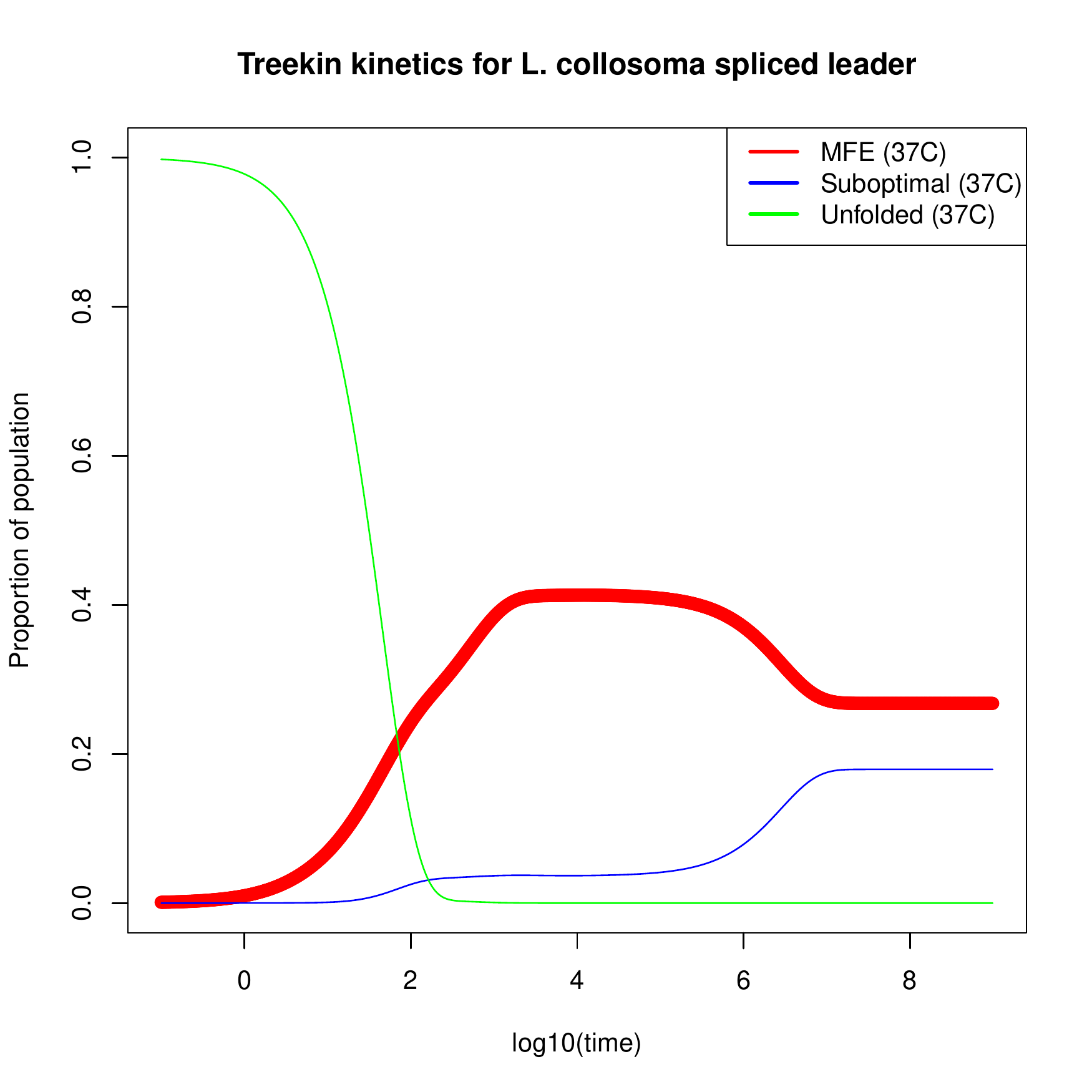}
\caption{ 
{\em (Left)}
Population occupancy curves computed with {\tt FFTeq} for the
56 nt conformational switch {\em L. collosoma} spliced leader RNA,
with sequence AACUAAAACA AUUUUUGAAG AACAGUUUCU GUACUUCAUU GGUAUGUAGA
GACUUC. The dot bracket format for the MFE structure, as computed by
version 2.1.7 of {\tt RNAfold -d0}, is {\tt
.......................((((((((((((.....)))))..)))))))..} with free
energy $-8.6$ kcal/mol, while that of the the alternate suboptimal
structure is {\tt
..((...((((((..(((((.((((...)))).)))))..))).)))..)).....} with free
energy $-7.5$ kcal/mol. In the case of the MFE structure, the
equilibrium occupancy $P(t_{\infty})$, which {\tt Hermes} approximates
as $0.17893806$ should equal the Boltzmann probability $0.17909227$,
since the MFE structure is the only structure at distance $x_0$ [resp.
$y_0$] from the reference structures $A$ (empty
structure) [resp. $B$ (MFE structure)]. As well, if there are few
other low energy structures at the same base pair distance $x_1$
[resp. $y_1$] from $A$ [resp. $B$] as that of the alternate suboptimal
structure, then we expect that the occupancy probability $0.03003426$
for the $(x_1,y_1)$ be approximately the Boltzmann probability
$0.03005854$ of the alternate structure. 
{\em (Right)} Population occupancy curves computed 
for {\em L. collosoma} spliced leader RNA, computed using
RNA Vienna Package with the following commands:
{\tt RNAsubopt -d0 -s -e 17  > foo.sub},
{\tt barriers -M noShift --max=1000 --rates --minh 0.1 < foo.sub  > foo.bar},
{\tt treekin --method I  --p0 940=1 --ratesfile rates.out  < foo.bar > out.bar},
where the value 940 indicates the empty structure appearing as the 940\textsuperscript{th}
structure in the sorted file {\tt foo.bar} of 1,000 locally optimal structures.
%{\em (Right)} Population occupancy curves computed 
%for {\em L. collosoma} spliced leader RNA, computed using
%RNA Vienna Package {\tt barriers} server
%{\tt http://rna.tbi.univie.ac.at/cgi-bin/barriers.cgi}, where 1000
%locally optimal structures were requested. After manually determining
%that the empty structure is in position 206 of the output from
%{\tt RNAsubopt}, 206 is entered as the initial structure whose population 
%density is set to 100\%, and the figure depicted here is generated.
The red curve corresponds to the {\em basin of attraction} around the
minimum free energy structure. At equilibrium, the Boltzmann probability
of this basin is between 0.3 and 0.4, a value substantially greater than
the Boltzmann probability of the minimum free energy structure itself,
which is $\approx 0.1791$ as shown in the left panel.}
\label{fig:populationOccupancySplicedLeader}
\end{figure*}

Since the state-of-the-art software {\tt Kinfold}, which implements
Gillespie's algorithm, cannot return statistically significant folding
times unless significant computation time is allotted, it is natural
to turn to {\em coarse-grained} models, as done by
Wolfinger et al. \cite{wolfingerStadler:kinetics} and by
Tang et al. \cite{Tang.jmb08}.  The software of Tang et al. appears not
to be available. Concerning the 
method of Wolfinger et al. (called {\tt BarrierBasins} in our benchmarking),
there is now a web server available, which runs {\tt RNAsubopt}
\cite{Wuchty.b99} to generate all secondary structures within a user-specified
energy range, then runs {\tt barriers} \cite{flammHofacker} to generate
basins of attraction around a user-specified number of locally optimal
structures, and then runs {\tt treekin} on the output of {\tt barriers}.
The program {\tt treekin} performs some of the same operations as {\tt Hermes},
by computing population occupation frequencies by spectral decomposition.
Nevertheless it would require a user to write scripts and perform 
several manual steps, in order to determine the equilibrium time for
an input RNA sequence, with respect to the macrostate Markov process
of \cite{wolfingerStadler:kinetics}. In addition, because {\tt barriers} computes
basins of attraction by utilizing the output of {\tt RNAsubopt}, estimating kinetics for the
refolding of an RNA molecule from the empty structure requires exhaustive enumeration
of all suboptimal structures having non-positive free energy.
Figure~\ref{fig:populationOccupancySplicedLeader} depicts some population
occupancy curves for the 56 nt spliced leader RNA from
{\em L. collorosoma}, a known conformational switch experimentally
investigated in \cite{lecuyerCrothers}, a determined by
{\tt Hermes} (left panel) and {\tt barriers} with {\tt treekin}
(right panel).

In contrast, {\tt Hermes} is a fully automated software suite which
computes mean first passage time and equilibrium time with respect to
the coarse-grained model that consists of macrostates $M(x,y)$ of all
structures having base pair distance $x$ to initial structure $A$ and
$y$ to target structure $B$. Additionally, {\tt Hermes} provides
a number of other functionalities useful for RNA kinetics. Since
{\tt Hermes} includes {\tt FFTbor2D}, which uses the fast Fourier transform
to compute probabilities $p(x,y)$ of macrostates $M(x,y)$, the resulting
coarse-grained kinetics computations are extremely fast. {\tt Hermes}
can provide approximate refolding kinetics for RNA sequences that are
larger than those which other software can handle. As well, the
correlation values shown in Table~\ref{table:correlation}
demonstrate that {\tt Hermes} can be used for fast, approximate and relatively
accurate RNA kinetics in a pipeline to select promising candidates in
the design of RNA sequences.

Finally, we mention an application of {\tt Hermes} to recent data from
Dotu et al. \cite{syntheticHammerheads}, which describes the
first purely computational design of functional synthetic hammerhead type III
ribozymes, experimentally shown to cleave {\em in vitro}. For the tiny
data set of 10 synthetic ribozymes, we ran {\tt FFTmfpt} to determine the
log(MFPT) of folding from the empty structure to the target consensus
structure of Peach Mosaic Latent Viroid (PMLVd), 
shown in Figure~\ref{fig:PLMVd}. We determined a
Pearson correlation of $0.4796$ [resp. $0.5322$]  between the logarithm of the 
{\em mean first passage time} [resp. {\em equilibrium time}], as computed by
{\tt FFTmfpt} [resp. {\tt RNAeq}] and the cleavage rate, using the
experimentally measured cleavage rate of 10 synthetic hammerhead ribozymes
(data from \cite{syntheticHammerheads}).
These correlation values seem surprising, since we had expected that
rapidly folding hammerheads (low values of MFPT and ET) might be 
associated with fast cleavage. The correlation values suggest instead
that perhaps the longer a hammerhead requires to fold into its functional
secondary structure, shown in Figure~\ref{fig:PLMVd}, the more rapid its
cleavage. If this observation can be experimentally validated on a large set 
of hammerheads, then the insight could be potentially important for
RNA design, an area of synthetic biology.

\section{Acknowledgements}

We would like to thank Ivan Dotu, for the suggestion of checking the
folding kinetics of the synthetic hammerheads in \cite{syntheticHammerheads}.
Some of the work described in this paper was done while PC visited
Niles Pierce at the California Institute of Technology, Knut Reinert
at the Free University of Berlin and Martin Vingron at the Max Planck
Institute for Molecular Genetics. Warm thanks are due to all three
persons. The research was funded by National Science Foundation grant
DBI-1262439, the Guggenheim Foundation and Deutscher Akademischer
Austauschdienst. Any opinions, findings, and conclusions or
recommendations expressed in this material are those of the authors
and do not necessarily reflect the views of the National Science
Foundation.

\clearpage
%\hfill\break\newpage
\section*{Appendix} 
\label{section:appendix}

\subsection*{Kinetics methods in benchmarking}

In this section, we describe the kinetics algorithms used in our
comparative study on a benchmarking set of 1000 small RNAs. The
kinetics algorithms we consider are the following: {\tt MFPT}, {\tt
Equilibrium}, {\tt Kinfold} \cite{flamm}, {\tt FFTmfpt}, {\tt FFTeq},
{\tt RNA2Dfold}, {\tt FFTbor1D}, {\tt BarrierBasins}
\cite{wolfingerStadler:kinetics}, {\tt Kinwalker} \cite{Geis.jmb08}.
Our software suite {\tt Hermes} introduces the new methods {\tt MFPT},
{\tt Equilibrium}, {\tt FFTmfpt}, {\tt FFTeq} (along with {\tt
RNA2Dfold} and {\tt FFTbor1D} through the utility programs {\tt
RNAmfpt} and {\tt RateEq}). Although all these methods allow one to
compute folding kinetics from an arbitrary initial structure to an
arbitrary final structure, in the benchmarking comparison, we consider
only folding from the empty structure to the MFE structure.
\begin{description}
\item[MFPT]: Given RNA sequence $\seq$ and secondary structure $x$,
let $N(x)$ denote the set of neighboring secondary structures of
$\seq$, whose base pair distance with $x$ is one. Define the Markov
chain $\mathcal{M}(\seq)=(SS(\seq),P)$, where $SS(\seq)$ denotes the
set of all secondary structures of $\seq$, $p^*(x)$ is the stationary
probability of structure $x$, defined by $p^*(x)=\exp(-E(x)/RT)/Z$,
$Z=\sum_{x} \exp(-E(x)/RT)$, and the transition probability matrix $P
= ( p_{x,y} )$ is defined either with or without the Hastings
modification as follows.

{\em With} the Hastings modification,
\begin{eqnarray}
\label{eqn:MFPTwithHastings} p_{x,y} = \left\{
\begin{array}{ll}
\frac{1}{|N(x)|} \cdot \min(1, \frac{p^*(y)}{p^*(x)} \cdot
\frac{N(x)}{N(y)}) &\mbox{if $y \in N(x)$}\\
0 &\mbox{if $x \ne y, y \not\in N(x)$}\\
1 - \sum_{\substack{z \in N(x)}} p_{x,z} &\mbox{if $x=y$}\\
\end{array}
\right.
\end{eqnarray}

{\em Without} the Hastings modification,
\begin{eqnarray}
\label{eqn:MFPTwithoutHastings} p_{x,y} = \left\{
\begin{array}{ll}
\frac{1}{|N(x)|} \cdot \min(1, \frac{p^*(y)}{p^*(x)} ) &\mbox{if $y
\in N(x)$}\\
0 &\mbox{if $x \ne y, y \not\in N(x)$}\\
1 - \sum_{\substack{z \in N(z)}} p_{x,z} &\mbox{if $x=y$}\\
\end{array}
\right.
\end{eqnarray}
The exact value of mean first passage time (MFPT) can be computed as
follows. Let $x_0$ [resp. $x_{\infty}$] denote the empty structure
[resp. MFE structure] for sequence $\seq$ (here we have implicitly
identified integer indices with secondary structures). Let
$\widetilde{P}_{x_{\infty}}$ be the matrix obtained from $P$ by
removal of the row and column with index $x_{\infty}$, and $I$ denote
the $(n-1)\times(n-1)$ identity matrix, where $n=|SS(\seq)|$ is the
number of secondary structures of $\seq$. Let $e$ denote the vector of
size $n-1$, each of whose coordinates is $1$. It is well-known
\cite{meyerMFPT} that for each $k\ne x_{\infty}$, the $k$th coordinate
of the vector $(I - \widetilde{P}_{x_{\infty}})^{-1} \cdot {\bf e}$ is
exactly equal to the mean first passage time from the structure with
index $k$ to the target structure $x_{\infty}$. In particular, the
MFPT from the empty structure to the MFE structure is computable by
applying matrix inversion from GSL. Since this computation of the
mean first passage time is mathematically exact, we consider that MFPT
to be the {\em gold standard} value for RNA folding kinetics.

\item[Equilibrium]: Define the continuous Markov process
$\mathcal{M}=(SS(\seq),R)$, where $R$ is the {\em rate matrix} defined
by
\begin{eqnarray}
\label{eqn:rateMatrix} k_{x,y} = \left\{
\begin{array}{ll}

%\frac{1}{|N(x)|} \cdot \min(1, \frac{p^*(y)}{p^*(x)} )
\min(1, \frac{p^*(y)}{p^*(x)} ) &\mbox{if $y \in N(x)$}\\
0 &\mbox{if $x \ne y, y \not\in N(x)$}\\
- \sum_{\substack{z \in N(z)}} p_{x,z} &\mbox{if $x=y$}\\
\end{array}
\right.
\end{eqnarray}
Clearly the rate matrix satisfies {\em detailed balance}; i.e. $p^*(x)
\cdot k_{x,y} = p^*(y) \cdot k_{y,x}$ for all distinct $x,y \in
SS(\seq)$. In fact, the rate matrix for Markov processes is usually
defined as above, precisely to ensure detailed balance, which then
implies that all eigenvalues of the rate matrix are real, thus
permitting explicit solution of the population occupancy frequency for
all states.\footnote{We additionally considered a Hastings correction
for the rate matrix, where $k_{x,y} = \min(1, \frac{p^*(y)}{p^*(x)}
\cdot \frac{N(x)}{N(y)})$. The correlation in
Table~\ref{table:correlation} for equilibrium time computed from this
modified rate matrix is somewhat better than without the Hastings
correction. However, the Hastings correction is never used for rate
matrices, hence we only consider the usual definition of rate matrix
given in equation (\ref{eqn:rateMatrix}).}

\item[Kinfold]: Version 1.3 of {\tt Kinfold} \cite{Flamm:00a}, from
the Vienna RNA Package (version 2.1.7) was run with the following
flags:
\begin{quote}
{\tt Kinfold --dangle 0 --met --noShift --logML --num=10000
--time=100000000}

%< infile.txt >\& /dev/null
\end{quote}
{\tt Kinfold} 1.3 uses the Turner2004 energy model, and the flags in
the command line indicate that dangles are not considered, that the
Metropolis rule is employed (rather than the Kawasaki rule), that
moves consist only of the addition or removal of a single base pair,
that the number of repeated Monte Carlo runs per sequence is $10^4$,
and that each run has an upper bound of $10^8$ steps. For the RNA
sequences of 20 nt used in our benchmarking set, with these flags,
every run for every sequence properly terminated in the target minimum
free energy structure. For each sequence, the mean (and standard
deviation) of $10^4$ runs was determined; this mean is deemed the mean
first passage time, as estimated by {\tt Kinfold}. Implemented in the
Vienna RNA Package 2.1.7, {\tt Kinfold} is what may be considered the
{\em silver standard} value for folding kinetics, since exact
determination of mean first passage time, using matrix inversion is
impossible except for sequences not having more than several thousand
secondary structures. However, due to the stochastic nature of the
Gillespie algorithm implemented by {\tt Kinfold}, very many runs with
very long run times are required to obtain a reasonable approximation
of the MFPT.

\item[FFTmfpt]: Given an RNA sequence $\seq$, the algorithm {\tt
FFTbor2D} \cite{Senter.jmb14} computes the probabilities $p(x,y) =
\frac{Z(x,y)}{Z}$ of secondary structures of $\seq$ to have base pair
distance $x$ [resp. $y$] to secondary structure $A$ [resp. $B$], where
$A$ is taken to be the empty structure and $B$ is the MFE structure.
Let $d_{BP}(S,T)$ denote the base pair distance between structures
$S,T$. Then $Z(x,y) = \sum_S \exp(-E(S)/RT)$, where the sum is over
structures $S$, such that $d_{BP}(A,S)=x$, and $d_{BP}(B,S)=y$; as
well the partition function $Z =\sum_S \exp(-E(S)/RT)$, where the sum
is over all seconddary structures $S$ of the sequence $\seq$.

Let $d_0= d_{BP}(A,B)$, the base pair distance between initial structure
$A$ and target structure $B$.  Let $n$ denote the length of sequence $\seq$.
Define the Markov chain $\mathcal{M}_1(\seq) = (Q_1,P_1)$, where
\begin{eqnarray}
\label{eqn:defQ1} Q_1 &=& \{ (x,y) : 0 \leq x,y \leq n, (x+y \bmod 2)
= (d_0 \bmod 2), \\ % Parity condition
& & (d_0 \leq x+y), (x \leq d_0 + y), (y \leq d_0 + x). \nonumber %
Triangle inequality
\end{eqnarray}
For reference, we say that the {\em parity condition} holds for
$(x,y)$ if
\begin{eqnarray}
\label{eqn:parityCondition} (x+y \bmod 2) = (d_0 \bmod 2).
\end{eqnarray}
We say that the {\em triangle inequality} holds for $(x,y)$ if
\begin{eqnarray}
\label{eqn:triangleInequality} (d_0 \leq x+y), (x \leq d_0 + y), (y
\leq d_0 + x)
\end{eqnarray}
Figure~\ref{fig:checkerboardGridPointsSplicedLeader} displays the
space of possible grid points $(x,y)$ for spliced leader RNA from {\em
L. collosoma} with two reference structures.

Since we allow transitions between secondary structures that differ by
exactly one base pair, Markov chain transitions are allowed to occur
only between states $(x,y),(u,v) \in Q_1$, such that $u=x\pm 1$, $v =
y \pm 1$. However, we have found that for some RNA sequences, there is
no non-zero probability path from $(0,d_0)$ to $(d_0,0)$
(corresponding to a folding pathway from structure $A$ to
$B$).\footnote{Since {\tt FFTbor2D} computes probabilities $p(x,y)$ by
polynomial interpolation using the fast Fourier transform, any
probability less than $10^{-8}$ is set to $0$. Also with {\tt
RNA2Dfold}, it may arise that there is no non-zero probability path
from structure $(0,d_0)$ to $(d_0,0)$.}

To address this situation, we proceed as follows. Let $\epsilon =
10^{-8}$ and for all $(x,y) \in Q_1$, modify probabilities $p(x,y)$ by
\begin{eqnarray}
\label{eqn:modifyNormalizeCellProb} p(x,y) &=& \frac{p(x,y) +
\epsilon/|Q_1|}{1+\epsilon }.
\end{eqnarray}
This operation corresponds to adding the negligeable value of
$\epsilon=10^{-8}$ to the total probability, thus ensuring that there
are paths of non-zero probability between any two states. After this
$\epsilon$-modification and renormalization, {\em when using the
Hastings modification}, the transition probabilities $P( (u,v) | (x,y)
)$ are given by
\begin{eqnarray}
\label{eqn:transitionProbabilityFFTbor2DwithHastings} P( (u,v) |(x,y)
) = \left\{
\begin{array}{ll}
\frac{1}{|N(x,y)|} \cdot \min(1, \frac{p(u,v)}{p(x,y)} \cdot
\frac{N(x,y)}{N(u,v)}) &\mbox{if $(u,v) \in N(x,y)$}\\
0 &\mbox{if $(x,y) \ne (u,v) \mbox{ and } (u,v) \not\in N(x,y)$}\\
1 - \sum_{\substack{(u,v) \in N(x,y)}} P( (u,v) | (x,y) ) &\mbox{if
$(x,y) =(u,v)$}\\
\end{array}
\right.
\end{eqnarray}
Here the set $N(x,y)$ of adjacent neighbors is defined by $N(x,y) = \{
(u,v) \in Q_1 : u = x \pm 1, v = y \pm 1 \}$, and the stationary
probability $p(x,y)$ is obtained from {\tt FFTbor2D}.

{\em Without the Hastings modification}, the transition probabilities
$P( (u,v) | (x,y) )$ are instead given by
\begin{eqnarray}
\label{eqn:transitionProbabilityFFTbor2DwithoutHastings} P( (u,v)
|(x,y) ) = \left\{
\begin{array}{ll}
\frac{1}{|N(x,y)|} \cdot \min(1, \frac{p(u,v)}{p(x,y)}) &\mbox{if
$(u,v) \ne (x,y)$}\\
1 - \sum_{\substack{(u,v) \ne (x,y)}} P( (u,v) | (x,y) ) &\mbox{if
$(x,y) =(u,v)$}\\
\end{array}
\right.
\end{eqnarray}

\item[FFTeq]: This method consists of computing the equilibrium time
from the master equation for the coarse-grain Markov process
$\mathcal{M}=(Q_1,R)$, where $Q_1$ is defined in equation
(\ref{eqn:defQ1}), and the rate matrix $R = ( k( (x,y), (u,v) ) )$ is
defined by
\begin{eqnarray}
\label{eqn:rateProbabilityFFTbor2DwithoutHastings} k( (x,y), (u,v) ) =
\left\{
\begin{array}{ll}
\min(1, \frac{p(u,v)}{p(x,y)}) &\mbox{if $(u,v) \ne (x,y)$}\\
1 - \sum_{\substack{(u,v) \ne (x,y)}} P( (u,v) | (x,y) ) &\mbox{if
$(x,y) =(u,v)$}\\
\end{array}
\right.
\end{eqnarray}
Equilibrium time is then computed for this Markov process.

\item[RNA2Dfold]: The program {\tt RNA2Dfold} of Vienna RNA Package
2.1.7 was run to compute the probabilities $p(x,y) = \frac{Z(x,y)}{Z}$
of secondary structures of $\seq$ to have base pair distance $x$
[resp. $y$] to secondary structure $A$ [resp. $B$], where $A$ is the
empty structure and $B$ is the minimum free energy structure. As in
the case for {\tt FFTbor2D},
Figure~\ref{fig:checkerboardGridPointsSplicedLeader} displays the
checkerboard image of possible grid points $(x,y)$ for spliced leader
RNA from {\em L. collosoma} with two reference structures.

The probabilities $p(x,y)$ were then $\epsilon$-modified and
normalized, as in the description above for {\tt FFTeq}, and the mean
first passage time for the Markov chain $\mathcal{M}_2(\seq) =
(Q_1,P_2)$ is computed, where $P_2$ is the transition probability
matrix, as in {\tt FFTbor2D}, with the exception that $p(x,y)$ is
initially obtained from {\tt RNA2Dfold}, rather than {\tt FFTbor2D}.

When working with {\tt
RNA2Dfold}, it is more accurate to compute transition probabilities of
the form $$ P( (u,v) | (x,y) ) = \frac{1}{N(x,y)} \cdot \min(1,
\exp(-\frac{E(u,v)-E(x,y)}{RT})) $$ instead of the mathematically
equivalent form $$ P( (u,v) | (x,y) ) = \frac{1}{N(x,y)} \cdot \min(1,
\frac{p(u,v)}{p(x,y)}) $$ for reasons due solely to numerical
precision. However, since it is necessary to perform
$\epsilon$-modification and renormalization of transition
probabilities,\footnote{Otherwise, as earlier explained, there may be
no non-zero probability path from $(0,d_0)$ to $(d_0,0)$.} one is led
to modify the free energy in the following manner.

Letting $G$ denote ensemble free energy $-RT \ln Z$, as computed by
{\tt RNAfold} and {\tt RNA2Dfold},
\begin{eqnarray*}
p &=& \exp(-E/RT)/\exp(-G/RT) \\
&=& \exp(-[ E-G] /RT)
\end{eqnarray*}
hence it follows that $-RT \ln p = E-G$ and $E = G-RT \ln p$.

By adding $\epsilon=10^{-8}/|Q_1|$ to every probability in every state
$(x,y) \in Q_1$, the $\epsilon$-modified and renormalized probability
corresponding to $p$ is $\frac{p+\epsilon/|Q_1|}{1+ \epsilon}$ and so
the corresponding new energy term is
\begin{eqnarray}
\label{eqn:modifiedEnergyTerm} Enew &=& G- RT \ln( \frac{p+\epsilon}{1
+ \epsilon \cdot |Q_1|} ).
\end{eqnarray}
It follows that we are forced to use probabilities in any case, and so
we encounter the above-mentioned loss of accuracy.

\item[BarrierBasins]: Using the {\tt --rate} flag for the program {\tt
barriers} \cite{flammHofacker}, Wolfinger et al.
\cite{wolfingerStadler:kinetics} computed the rate matrix for a Markov
process, whose states are the basins of attraction around locally
optimal secondary structures. Subsequently, using {\tt Treekin}, the
authors displayed population occupancy curves for certain macrostates
(basins of attraction) containing specific structures of interest.
Since this method was not given a name by the authors, we let {\tt
BarrierBasins} denote the equilibrium time for the Markov process just
described.

\item[Kinwalker]: {\tt Kinwalker} was used with default settings.
However, since {\tt Kinwalker} models co-transcriptional folding of
RNA, its correlation was extremely poor with all other methods, which
do model refolding, rather than co-transcriptional folding (data not
shown). For this reason, we do not display the results with {\tt
Kinwalker}.
\end{description}

\bibliographystyle{plain}
%\bibliography{/Users/clote/text/BIBdir/clote}
%\bibliography{biblio}

\end{document}